\documentclass{article}

\usepackage{microtype}
\usepackage{graphicx}
\usepackage{subfigure}
\usepackage{booktabs} 
\usepackage{stfloats}
\usepackage{mathrsfs}

\usepackage{hyperref}

\usepackage[accepted]{icml2025}

\usepackage{amsmath}
\usepackage{amssymb}
\usepackage{mathtools}
\usepackage{amsthm}

\usepackage[capitalize,noabbrev]{cleveref}

\theoremstyle{plain}

\theoremstyle{definition}

\theoremstyle{remark}

\usepackage[textsize=tiny]{todonotes}

\icmltitlerunning{Construct to Commitment}

\begin{document}

\twocolumn[
\icmltitle{Construct to Commitment: The Effect of Narratives on Economic Growth}



\icmlsetsymbol{equal}{*}

\begin{icmlauthorlist}
\icmlauthor{Hanyuan Jiang}{equal,yyy}
\icmlauthor{Yi Man}{equal,sch}

\end{icmlauthorlist}

\icmlaffiliation{yyy}{Department of Economics, London School of Economics and Political Science, Houghton Street, London WC2A 2AE, UK}
\icmlaffiliation{sch}{School of International Studies, Zhejiang University, Yuhangtang Road, Hangzhou 310058, China}

\icmlcorrespondingauthor{Hanyuan Jiang}{h.jiang32@lse.ac.uk}
\icmlcorrespondingauthor{Yi Man}{3220101358@zju.edu.cn}

\icmlkeywords{Algorithmic game theory, Dynamic Contract Theory, Machine Learning, Macroeconomics, Political Economics, Communication Theory}

\vskip 0.3in
]



\printAffiliationsAndNotice{\icmlEqualContribution} 

\begin{abstract}
We study how government-led narratives through mass media evolve from construct, a mechanism for framing expectations, into commitment, a sustainable pillar for growth. We propose the ``Narratives-Construct-Commitment (NCC)" framework outlining the mechanism and institutionalization of narratives, and formalize it as a dynamic Bayesian game. Using the Innovation-Driven Development Strategy (2016) as a case study, we identify the narrative shock from high-frequency financial data and trace its impact using local projection method. By shaping expectations, credible narratives institutionalize investment incentives, channel resources into R\&D, and facilitate sustained improvements in total factor productivity (TFP). Our findings strive to provide insights into the New Quality Productive Forces initiative, highlighting the role of narratives in transforming vision into tangible economic growth.
\end{abstract}

\section{Introduction}
\pdfbookmark[1]{Introduction}{sec:introduction}

We study issues in Chinese path to modernization. China's economy has transitioned from a phase of high-speed growth to high-quality development. Socioeconomic development requires policy guidance, where governments play a pivotal role in steering economic direction, optimizing economic structures, and promoting industrial upgrading through the formulation and implementation of effective policies. Academic research typically concentrates on ex-ante policy simulations (e.g., DSGE models) or ex-post policy evaluations (e.g., Difference-in-Differences [DID], Regression Discontinuity Design [RDD]). However, the complex and dynamic interaction mechanisms among central government intentions, local government implementation, and market expectations during policy execution remain underexplored. For instance, despite central industrial policies explicitly emphasizing industrial upgrading and capacity optimization, why do certain regions still struggle with persistent overcapacity and structural supply-demand mismatches? We argue that the crux lies not in the optimality of policy design, but in how policy perceptions are gradually shaped, implemented, disseminated, and internalized. In other words, policy effectiveness depends not only on its \textit{de jure} design but crucially also on its \textit{de facto} interpretation, internalization, and eventual transformation into socially recognized long-term commitments.

We propose the ``Narrative-Construct-Commitment" (NCC) framework, emphasizing that policy implementation constitutes not a discrete event but a continuous dynamic process of communication and cognition. Unlike traditional economic approaches (e.g., DSGE, DID, RDD) that adopt either ex-ante or ex-post perspectives, the NCC framework examines the or entire dynamic progression from initial narrative articulation to the endogenous transformation of policies into long-term growth drivers from a mechanism design perspective. Conventional methods inadequately capture critical elements such as signal transmission, cognitive biases, and expectation management during policy implementation, necessitating a more dynamic, process-oriented methodological complement.

Specifically, ex-ante approaches like Dynamic Stochastic General Equilibrium (DSGE) models, while providing theoretical predictions under strict rational expectations assumptions, insufficiently address real-world policy complexities, particularly cognitive and decision-making biases under bounded rationality (Gabaix 2020). Recent advancements (Farhi \& Werning 2019) incorporate concepts like "cognitive discounting"where agents exhibit "partial myopia" when evaluating future eventsto better explain macroeconomic phenomena such as forward guidance puzzles and zero lower bound constraints. However, the nature, origins, and formation mechanisms of cognitive discounting, along with its integration with other behavioral biases, including risk perception, uncertainty aversion, information asymmetry, limited attention, anchoring effects, framing effects (Kahneman 1973), remain understudied. Furthermore, incorporating behavioral mechanisms increases model complexity without resolving the limitations of linear approximation methods typically employed in model solutions.

Ex-post evaluation methods, including DID and RDD, quantitatively assess policy impacts through real-world data, offering robust causal inference (Imbens \& Lemieux 2008). Nevertheless, these approaches predominantly focus on final outcomes while neglecting dynamic mechanisms and implementation processes. For example, although DID might identify positive effects of new energy subsidies on corporate innovation (Acemoglu et al. 2019), it fails to analyze deviations in local government interpretations, how policy narratives gradually shape market expectations, or how firms internalize beliefs about policy credibilityall critical to understanding policy effectiveness.

To bridge this gap, our NCC framework focuses on the interim implementation process, conceptualizing policy execution as a dynamic continuum of narrative dissemination and cognitive internalization. Integrating ex-ante attention to expectation formation with ex-post emphasis on causal effects, this framework reveals how policy narratives influence cognitive frameworks of market participants, thereby achieving sustainable long-term growth. The NCC framework systematically examines three dimensions: Narrative, Construct, and Commitment, respectively delineating how policy narratives systematically shape market expectations; how initial policy dissemination builds social cognitive consensus; and how narratives evolve into stable beliefs and enduring growth drivers through prolonged practice. This approach enables analysis of transmission and cognitive challenges during policy implementation while providing governments with a dynamic, process-oriented tool to enhance narrative-to-execution effectiveness.

Using the Innovation-Driven Development Strategy (IDDS) officially issued by the State Council in 2016 as an empirical case, we employ high-frequency financial market data and local projection to analyze how initial policy narratives anchor market expectations (Construct) and assess their long-term transformation into credible commitments (Commitment). Our study not only validates the NCC framework but also strive to offer practical insights for the New Quality Productive Forces (NQPFs) initiative. By elucidating how policy narratives influence macroeconomic outcomes, shape long-term corporate behavior, and guide investment decisions, this research advances theoretical understanding of Chinese-style modernization while informing global policy practice.

\section{Narratives-Construct-Commitment: An Overview}
\pdfbookmark[1]{Narratives-Construct-Commitment: An Overview}{sec:intro}

In this part we discuss the main theory. Drawing upon inspirations from economics, communication studies, and political science, the Narratives-Construct-Commitment (NCC) framework constitutes a complete chain from policy narrative initiation to institutionalization, reflecting the dynamic process of government policies from vision building to policy implementation. \textbf{Narratives} serve as the foundational starting point of policy transmission mechanisms, referring to the comprehensive policy discourse system constructed by governments through language and rhetoric. Such narratives not only provide contextual background and visions for policy objectives but also establish the basis for societal dissemination. Though narratives alone do not directly translate into policy implementation or social action, they shape a broad cognitive context that determines whether policies can initially attract widespread attention and social resonance (Shiller 2017). For instance, China's reform and opening-up as a grand creation from theory to practice leveraged systematic policy narratives ranging from ``practice is the sole criterion for testing truth" to ``crossing the river by feeling the stones." These narratives not only fostered societal consensus on market-oriented reforms but also evolved into institutional commitments that propelled China's sustained high-speed economic growth.

However, macro-level narratives alone cannot effectively guide specific market expectations and societal cognition. Thus, we introduce \textbf{Construct} as the critical bridge between narratives and concrete policy commitments. Unlike broad narratives, Construct emphasizes the systematic and structured refinement of policy narratives, requiring governments to translate abstract visions into specific and well-defined social expectations and cognitive frameworks through carefully designed linguistic and communicative strategies. Construct does not represent final policy commitments but rather the preliminary process of anchoring expectationstransforming abstract visions into actionable directions and psychological foundations for societal cognition. The effectiveness of Construct determines whether policy narratives can establish broad and stable belief systems within society, thereby influencing economic decisions and resource allocation among market participants.

When Construct successfully achieves widespread societal acceptance and internalization of policy narratives, policies may enter the \textbf{Commitment} phase. Distinct from the preliminary cognition of the Construct stage, Commitment refers to the institutionalization of policies through prolonged societal cognition and practical validation, forming stable social expectations and long-term policy constraints. At this stage, policy narratives no longer merely serve as cognitive tools but become institutionalized components of fiscal resource allocation, legal regulations, and industrial policy frameworks. Policy sustainability and credibility are ultimately realized during Commitment.

The relationships can be summarized as follows: \textbf{Narrative} provides the foundational framework for policy dissemination, offering macro-level visions and potential bases for societal consensus; \textbf{Construct} systematizes and structures narratives to shape and stabilize specific expectations and cognition among market participants, laying groundwork for policy execution; {\textbf{Commitment}} represents the final stage where policies, after long-term societal cognition and practical validation, become endogenous and institutionalized as sustainable tools for long-term economic growth.

\subsection{Narratives}
\pdfbookmark[2]{Narratives}{sec:intro}

Narratives have increasingly been recognized as a core concept for understanding economic behavior and policy impacts. Unlike traditional economic theories that emphasize fully rational decision-making with perfect information, we focus on how individuals make economic decisions under bounded rationality and incomplete information through story-based interpretive frameworks and shared collective beliefs. Narratives are not merely vehicles for information transmission but powerful cognitive-shaping tools that significantly influence public and market expectations and behavioral patterns.

Shiller (2017), in his \textit{Narrative Economics}, argues that major economic events and cyclical fluctuations are often driven by the rise, dissemination, and decline of specific narratives. By shaping public psychological expectations, narratives directly affect macroeconomic variables such as consumption, investment, and employment. This perspective fundamentally diverges from traditional rational expectation models, which assume full rationality among economic agents while neglecting the profound impact of social psychology and collective beliefs on economic fluctuations. Our formulation is also inspired by the Forward Guidance, which originates from central banks influencing market expectations and achieving macro goals through explicit future policy statements (Bernanke 2020). Evans' (2017) categorization of Forward Guidance into \textit{Odyssean Guidance} (explicit future policy commitments) and \textit{Delphic Guidance} (economic forecasts shaping expectations without binding commitments) further clarifies the distinction between Construct and Commitment in policy narratives.

Forward Guidance has become a critical tool for central banks in developed economies to communicate with and manage public expectations. Its efficacy lies not in directly altering real economic variables but in influencing market expectationsa process enabled by human bounded rationality (Simon 1957) and decision-making biases (Kahneman \& Tversky 1979). Prospect Theory demonstrates that individuals respond differently to policy goals depending on how they are framed, altering perceived risks and rewards (Kahneman \& Tversky 1981). Communication studies' Framing Theory (Entman 1993) similarly posits that the framing of narratives determines how information is interpreted. For instance, whether a policy is framed as emphasizing growth opportunities or risk mitigation significantly affects public acceptance and expectation formation. Individuals do not first perceive events and then apply frames; rather, they interpret information through preexisting cognitive frameworks until contradictions arise (McCombs 1997).

Kingdon's (1984) "policy window" theory highlights that policy narratives gain traction during specific historical or economic contexts. Hall's (1993) "policy paradigm" theory emphasizes that when narratives stabilize into dominant paradigms, government decisions, industrial structures, and resource allocation undergo transformative shifts. Thus, policy narratives are not merely short-term communication tools but mechanisms that evolve into institutionalized cognitive frameworks and long-term policy constraints. Agenda-Setting Theory (McCombs \& Shaw 1972) underscores the \textit{salience} of policy issueshow governments use media to prioritize topics and amplify their significance. Unlike framing effects, agenda-setting determines \textit{which} issues gain public attention rather than \textit{how} they are understood (Severin \& Tankard 2001). Construct integrates both framing and agenda-setting: policy narratives simultaneously determine which issues markets and the public prioritize \textit{and} how they interpret them.

In China's policy practice, narratives have been widely deployed in disseminating macroeconomic\&industrial policies, and national development initiatives. Concepts such as ``high-quality development," ``supply-side structural reform," and "innovation-driven development" were \textit{disseminated} through narratives during early policy formulation, successfully fostering market and societal understanding and alignment with policy objectives.

\subsection{Construct}
\pdfbookmark[2]{Construct}{sec:intro}

The term construct is widely used in academia to express the meaning of social construct, emphasizing that social facts do not exist naturally but are created through language, culture, and collective understanding (Berger \& Luckmann 1966). We explicitly define Construct as the initial building phase of policy narratives, a dynamic policy mechanism that systematically influences market and social cognition through narratives and discourse systems, laying the cognitive foundation for policy institutionalization and long-term economic growth (Commitment). Construct influences how market participants and the general public understand policies, thereby guiding long-term development expectations. It not only sets the direction of policies but also builds a belief in the policy vision, making it the cognitive basis for economic and social development. Construct aims to shape consensus, enhance trust, guide expectations, and is a key link in transforming policies from ideas to reality. Construct is similar to the previously mentioned Delphic Guidance, emphasizing the initial role of government policy narratives in shaping market expectations, cognitive frameworks, and investment decisions, but Construct covers a broader range of policies, not limited to monetary policy or financial markets.

In the development of socialism with Chinese characteristics, top-level design has effectively transformed policy concepts into social consensus through Construct. Construct combines policy objectives with social development realities, providing expectation guidance for economic development, technological innovation, and industrial upgrading, thereby enhancing policy implementation and credibility. We state that Construct has three main characteristics:

\begin{itemize}
  \item The core objective of Construct is to shape market participants' policy expectations, forming consensus among various sectors of society regarding policy direction. It relies on Policy Narratives and Framing to transform policy objectives into an understandable, acceptable, and anticipated social cognitive system. The effectiveness of Construct depends on the systematization of policy narratives, the clarity of discourse systems, and the consistency of policy signals. The government needs to ensure the coherence and comprehensibility of policy information, transforming macro and abstract policy narratives into more specific, perceptible policy signals, enabling market participants to accurately anticipate policy directions and optimize economic behaviors such as investment, production, and consumption;
  \item Construct is not a static policy tool but a dynamically evolving communication mechanism. It continuously reinforces policy objectives, signals, and directions through a multi-level, multi-dimensional policy communication system, making them part of the social cognitive system. The government forms stable social consensus by disseminating policies through official media, public communication channels, social research institutions, and emerging digital platforms, ensuring broad understanding across different social groups, and continuously adjusting and optimizing based on policy implementation;
  \item Construct embodies the advantages of the socialist system with Chinese characteristics and must be tailored to local conditions, closely integrated with socialist thought with Chinese characteristics, current economic conditions, and long-term development goals. While reflecting the advantages of the socialist system and the modernization of governance capabilities, it must meet the practical needs of economic and social development, ensuring that policy narratives have both practical relevance and long-term development guidance. This coordination and consistency effectively ensure that policy narratives not only guide market expectations effectively in the short term but also lay a solid social cognitive foundation for long-term policy implementation (Commitment).
\end{itemize} 

Imagine a simple and intuitive example: a government aims to promote the development of artificial intelligence (AI) industry. Rather than immediately introducing specific industrial support policies or clear fiscal incentive measures, the government first constructs a preliminary narrative framework through channels such as press briefings, ``Artificial intelligence is the core driving force leading a new round of technological revolution and industrial transformation," positioning it as a key area for national development. Subsequently,  media intensively report on global AI development trends, closely linking them with  manufacturing transformation, upgrading, and digital economy development; government work reports and initiatives list AI as a priority development area, and relevant ministries issue supportive policy documents; experts and scholars interpret policy documents, academic institutions and think tanks publish research reports analyzing the positive impact of AI on economic and social development. This series of coordinated and consistent narrative frameworks not only highlights the historical inevitability and strategic significance of developing AI but also clearly outlines the policy direction, guiding enterprises, capital, and talent to concentrate in the AI field, forming a reallocation of economic resources.

In this process, Construct transforms abstract policy objectives into specific expectations and action guidelines for market participants through systematic, multi-level narrative construction, thereby producing substantive economic impacts, prompting economic participants such as enterprises, financial institutions, and local governments to actively adjust their investment and decision-making behaviors and resource allocation even before seeing specific policy measures. At this point, although the specific details of the policy have not yet been introduced, the expectations of the public and market participants have initially been anchored, and social belief in policy objectives has begun to form gradually. It can be seen that Construct, as a narrative mechanism, first establishes consensus on the importance of policies at the cognitive level among markets and the public, and initially reduces the resistance and economic uncertainty that may be encountered during future policy implementation.

\subsection{Commitment}
\pdfbookmark[2]{Commitment}{sec:intro}

The concept of Commitment originates from economic research on policy credibility and time consistency (Kydland \& Prescott 1977). They pointed out that the lack of long-term policy commitment mechanisms by governments would lead to unstable public and market expectations, reduced policy implementation efficiency, and hindered economic growth. This paper further redefines Commitment as an endogenized social belief mechanism, which not only guarantees policy stability and continuity but also provides a sustained and stable source of power for long-term sustainable economic growth. Therefore, we believe that Commitment is not a single policy norm or constraint, but a sustainable mechanism that is deeply internalized as social consensus and ultimately manifested as the endogenous driving force for economic growth.

From the perspective of growth theory, Commitment is closely related to endogenous growth which emphasizes that sustained economic growth stems from technological innovation, human capital accumulation, and knowledge spillover effects within the economic system (Romer 1990; Lucas 1988). Through sustained and stable policy commitments, the government significantly reduces the uncertainty faced by enterprises in long-term innovation investment and human capital accumulation, thereby increasing the expected rate of return for economic participants and enhancing their willingness for innovation input and long-term investment (Acemoglu, Johnson, and Robinson 2005). We believe this mechanism reduces the negative impact of economic fluctuations on long-term investment decisions, enhances firms' belief in innovation returns, and subsequently promotes corporate R\&D investment and human capital input, thereby endogenously increasing the long-term productivity growth of the economy, successfully implementing policy narratives as growth commitments.

Continue with the example above, the government clearly proposes and gradually implements a long-term AI industry development strategy. After the effective dissemination of preliminary narratives in the Construct phase, the government further establishes clear fiscal policy support, legal guarantees, and long-term incentive mechanisms. For example, by setting up stable research funds, introducing industry preferential policies, improving intellectual property protection systems, and maintaining a relevant institutional environment for the long term. These measures gradually convince market participants that the development of the AI industry is not just a short-term focus of the government but a long-term and stable direction for economic growth, and policies will not fundamentally change due to short-term economic fluctuations or political changes.

As policies continue to be implemented steadily, markets and enterprises gradually internalize this policy cognition as part of their own development strategies. Based on this long-term belief, enterprises continuously expand R\&D investment, introduce and cultivate high-end talent, promote technological breakthroughs and innovation activities, thereby increasing the endogenous growth rate of At. Meanwhile, capital markets also continue to increase funding support for related fields due to long-term stable expectations of policies. Ultimately, this stable and endogenous social belief mechanism (Commitment) successfully transforms into a long-term driving force for sustainable economic growth. If Construct is a cognitive and expectation management mechanism in the initial stage of policy, with relatively short-term impact mainly limited to information dissemination and expectation guidance, then Commitment is an endogenized, stabilized belief mechanism formed after long-term verification and continuous practice of policy narratives, with more long-term and fundamental influence, penetrating into the internal economic production function and becoming an important contributor.

In short, Construct makes policies possible, while Commitment makes policies reality and transforms them into stable sources of endogenous growth. This transformation from exogenous policy concepts to endogenized beliefs is the key to economic policies moving from short-term expectation management to long-term  growth and high-quality development.

\section{The Model}
\pdfbookmark[1]{The Model}{sec:model}

 Section I qualitatively explored the dynamic interaction process between government in Narrative, Construct, and Commitment (NCC framework), analyzing the impact of information asymmetry between central and local governments, differences in local governance capabilities, and credibility accumulation on policy implementation effects. To further clarify the internal logic of these mechanisms and determine the optimal policy strategy combination through quantitative analysis, we introduce a recursive structure model that formalizes the NCC framework as a dynamic Bayesian game between central, local governments, and market participants. In this model we treat policy narrative precision and policy implementation consistency as two cost signals released by the government (dual signals), with market participants dynamically updating their perception of government credibility based on these signals, ultimately triggering institutional commitment.

Our model is inspired by Contract Theory. Similar to Spence's (1973) signaling model, we view the narrative precision ($p$) set by the government and the implementation consistency ($c$) of local governments as cost signals of government type (governance capability), helping market participants identify unobservable governance capabilities through observable signals. Referencing Akerlof (1970) and Stiglitz (1975)'s screening theory, we assume that market participants use Bayesian updating mechanisms to correct their beliefs of government credibility based on local governments' actual implementation ($\theta$), avoiding market failure caused by information asymmetry. Drawing on Baker et al. (2016)'s policy shock identification method, we examine the different impacts of narrative shocks (short-term policy fluctuations) and implementation shocks on long-term institutional lock-in in a dynamic context, thereby clarifying how the ``dual signal" mechanism promotes credibility accumulation and policy effectiveness.

\subsection{Setup}
\pdfbookmark[2]{Setup}{sec:model}

We consider an infinite-horizon dynamic environment denoted as $t = 0,1,2,...$. The economy consists of three types of agents: Central Government, Local Governments, and Market Participants (participants). \\

\textbf{Central Government.} Responsible for establishing and announcing the policy narrative, act as the incumbent. In each period, the central government determines two critical parameters:

\begin{itemize}
\item Narrative precision $p_t \in [0,1]$, indicating the clarity with which the central government conveys its policy objectives. A higher value implies that the central government explicitly articulates detailed and specific policy goals (typically containing quantified targets and concrete details). Conversely, a lower precision implies vagueness in policy descriptions. Producing a more precise narrative demands additional external resources, thus incurring a cost denoted by the function $C_{\text{central}}(p_t,m_t)$ (detailed explanations to follow).
\item Monitoring intensity $m_t \in [0,1]$, reflecting the central government's effort level in supervising local government policy execution, aimed at ensuring effective implementation. A higher monitoring intensity $m_t$ improves local governments' execution efficiency but simultaneously increases administrative and management costs.
\end{itemize}

The central government's utility function is:
\begin{equation*}
U_{\text{central},t} = \lambda I_{\text{total},t} - \gamma \text{Var}(I_{\text{total},t}) - C_{\text{central}}(p_t,m_t) - \psi L_t,
\end{equation*}
where $I_{\text{total},t}$ represents the investment level within the economy during period $t$, reflecting the market's confidence in policy implementation. Parameter $\lambda > 0$ denotes the central government's target weight assigned to economic growth. The parameter $\gamma > 0$ captures the central government's aversion to economic fluctuations; specifically, $\text{Var}(I_{\text{total},t})$ reflects investment volatility, serving as an indicator of negative economic impacts caused by policy uncertainty. $C_{\text{central}}(p_t,m_t)$ is the central government's cost in communicating precise narratives and implementing monitoring efforts. The degree of maintaining institutional order incurs cost $\psi L_t$, with $\psi > 0$.

\textbf{Local Governments.}Encompass multiple departments and administrative hierarchies involved in policy execution, characterized by complexity and multiplicity. For simplicity and clarity, our model synthesizes local governments into unified participants. Given disparities in economic development levels and governance capabilities, each local government faces heterogeneous execution costs represented by the intrinsic parameter $\kappa_{c,t}$. We assume:
\begin{equation*}
\kappa_{c,t}\sim U[\underline{\kappa},\overline{\kappa}],\quad 0<\underline{\kappa}<\overline{\kappa}<\infty.
\end{equation*}
Upon observing the central government's choices of $p_t$ and $m_t$, local governments decide their policy implementation consistency $c_t \in [0,1]$ based on the trade-off between resource constraints and local benefit considerations. The variable $c_t$ measures the degree to which local governments implement the policy. Specifically, $c_t \simeq p_t$ indicates high consistency in execution, whereas $c_t \ll p_t$ suggests incomplete policy implementation. Local governments incur execution costs determined by their intrinsic characteristics:
\begin{equation*}
C_{\text{local},t} = \frac{1}{2}\kappa_{c,t}c_t^2,
\end{equation*}

where $k_{c,t}$ measures the local government's administrative capacity. A lower capacity implies higher costs. Local governments choose $c_t$ to maximize their payoff function:
\begin{equation*}
U_{\text{local},t}=\xi\theta_t-\frac{1}{2}\kappa_{c,t}c_t^2-(\beta_0+m_t)(c_t-p_t)^2,
\end{equation*}
$\xi\theta_t$ denotes benefits gained by the local government from achieving higher credibility (the definition of $\theta_t$ will follow below). The term $(\beta_0+m_t)(c_t-p_t)^2$ represents penalties (such as accountability measures) resulting from deviation from the central narrative target $p_t$.

\textbf{Market participants.} Including enterprises and individualsdynamically update their beliefs regarding policy credibility $\theta_t\in[0,1]$ based on observed signals of narrative precision ($p_t$) and local execution consistency ($c_t$). Specifically, participants renew their belief about whether the government ``honors its promises" based on observed combinations $(p_t,c_t)$ each period. Following Baker et al. (2016) regarding the influence of policy-related information shocks on economic agents' expectations, we explicitly adopt a linear mapping and assume that market participants are risk-neutral (important for empirical analysis, see IV.A). Thus, agents' investment decisions depend on expected returns without incorporating additional risk premiums:
\begin{equation*}
\theta_{t+1}=\theta_t+\eta(c_tp_t-\theta_t)
\end{equation*}

Market participants are categorized into two groups:
\begin{itemize}
\item Believers (type B, fixed proportion $\alpha$): their investment directly correlates positively with narrative precision:
\begin{equation*}
I_{B,t}=k_Bp_t,\quad k_B>0.
\end{equation*}

\item Rational Participants (type R) and Skeptics (type S): their relative proportions evolve dynamically. Initially, the Skeptics' proportion is $\omega_t\in[0,1-\alpha]$, implying the Rational Participants' share is $1-\alpha-\omega_t$. Rational Participants update their investment levels through Bayesian updating based on $\theta_t$:
\begin{equation*}
I_{R,t}=(1-\alpha-\omega_t)k_R\theta_t,\quad k_R>0.
\end{equation*}

Skeptics, with proportion $\omega_t$, exhibit investment as:
\begin{equation*}
I_{S,t}=\omega_tk_S\overline{\theta},\quad k_S>0.
\end{equation*}
\end{itemize}

Here, $\overline{\theta}$ represents their constant belief, specifically ${E}[\theta_t|\mathcal{F}_{t-1}]$. Without loss of generality \footnote{See APPENDIX.B for justification.}, we assume $\overline{\theta}$ as the historical average belief (long-term steady-state credibility). Skeptics respond sluggishly to updated signals, and their proportion declines as consistency $c_t$ increases:
\begin{equation*}
\omega_{t+1}=\max\{\omega_t-n_Sc_t,0\},\quad n_S>0.
\end{equation*}

The market's aggregate investment scale is thus given by:
\begin{equation*}
\begin{split}
I_{\text{total},t} &= I_{B,t} + I_{R,t} + I_{S,t} \\
                   &= \alpha k_B p_t + (1-\alpha-\omega_t) k_R \theta_t + \omega_t k_S \overline{\theta}.
\end{split}
\end{equation*}

To capture the process whereby local governments internalize narrative credibility and gradually transition from Construct to Commitment, we introduce an internal state variable $L_t$, defined as the level of policy institutionalization, whose dynamics follow:
\begin{equation*}
L_{t+1}=L_t+\phi\theta_t\mathbf{1}\{\theta_t\geq\theta_{\text{threshold}}\},\quad\phi>0.
\end{equation*}

When $L_t\geq L_{\text{threshold}}$, local governments meet institutional commitment constraints:
\begin{equation*}c_t \geq c_{\text{min}}.
\end{equation*}

\subsection{Equilibrium Analysis and Optimal Strategies}
\pdfbookmark[2]{Equilibrium Analysis and Optimal Strategies}{sec:model}

Before proceeding with the analysis, we first introduce two key stochastic disturbances to characterize unobservable measurement errors or external shocks associated with the actual realization of policy signals and execution behavior:

\begin{itemize}
    \item Belief updating noise $\varepsilon_t \sim \mathcal{N}(0, \sigma_\varepsilon^2)$, representing disturbances in the market's updating process of credibility $\theta_t$;
    
    \item Execution noise $\nu_t \sim \mathcal{N}(0, \sigma_\nu^2)$, representing disturbances in the local government's realization of execution level $c_t$.
\end{itemize}

Under these stochastic conditions, the central government's dynamic planning problem requires an expectation-based formulation. Local governments, in turn, optimize their period utility, and market participants adjust their investment strategies accordingly. We proceed as follows.\\

\textbf{Local Governments.} Given the central government's predetermined narrative precision $p_t$ and monitoring intensity $m_t$, local governments' choice of execution consistency is subject to the stochastic shock $\nu_t$. Thus, the local government can only select an ideal execution level $c_t^*$, while the actual execution rule follows:
\begin{align*}
\label{eq:execution_rule}
c_t = c_t^* + \nu_t, \quad \nu_t \sim \mathcal{N}(0, \sigma_\nu^2).
\end{align*}

\textbf{Lemma 1:} Given central government strategies $(p_t, m_t)$, the local government's optimal ideal execution policy that maximizes expected period utility is:
\begin{align*}
c_t^* = \frac{(\beta_0 + m_t)p_t}{\kappa_{c,t} + (\beta_0 + m_t)}.
\end{align*}

Under institutional commitment constraints ($L_t \geq L_{\text{threshold}}$), the local government's decision rule becomes:

\begin{align*}
c_t^* = \max \left\{ \frac{(\beta_0 + m_t)p_t}{\kappa_{c,t} + (\beta_0 + m_t)}, \, c_{\text{min}} \right\}.
\end{align*}

\textbf{Proof:} Local governments' expected utility (see II.A) is:
\begin{align*}
{E}[U_{\text{local}}(c_t)] = \alpha \theta_t - \frac{1}{2} \kappa_{c,t} {E}[c_t^2] - (\beta_0 + m_t) {E}[(c_t - p_t)^2].
\end{align*}

Expanding the expectation and solving for the control variable $c_t^*$, we derive the first-order condition:
\begin{align*}
\frac{\partial {E}[U_{\text{local}}]}{\partial c_t^*} = -\kappa_{c,t} c_t^* - 2(\beta_0 + m_t)(c_t^* - p_t) = 0.
\end{align*}

Solving for $c_t^*$ yields the stated optimal solution. Incorporating the truncation condition, the lemma follows. QED.\\

\textbf{Market participants.} Observe the local governments' actual execution $c_t$ and update their beliefs regarding credibility $\theta_t$ as follows:
\begin{align*}
\theta_{t+1} = \theta_t + \eta (c_t p_t - \theta_t) + \varepsilon_t, \quad \varepsilon_t \sim \mathcal{N}(0, \sigma_\varepsilon^2).
\end{align*}

According to the setup in Section II.A, the proportion of Skeptics among market participants holds a constant historical belief $\overline{\theta}$ (long-term average credibility). This proportion $\omega_t$ decreases progressively as execution consistency $c_t$ improves:
\begin{align*}
\omega_{t+1} = \max \{ \omega_t - \eta_S c_t, 0 \}.
\end{align*}

Therefore, in the long run, market participants gradually transition from Skeptics to Rationals, reflecting the transition from the Construct phase to the Commitment phase.

The central government's dynamic planning problem is to maximize the expected discounted sum of intertemporal utility, with the objective function specified as:
\begin{align*}
\max_{\{p_t, m_t\}} \; E_0 \Bigg[
\sum_{t=0}^{\infty} \delta^t \Big(
    & \lambda I_{\text{total},t}
    - \gamma Var(I_{\text{total},t}) \\
    & - C_{\text{central}}(p_t, m_t)
    - \psi I_t
\Big)
\Bigg]
\end{align*}
where $I_{\text{total},t}$ denotes total investment scale in the market, as defined in Section II.A.\\

\textbf{Proposition 1} Existence of Central Government's Dynamic Planning Problem Solution: Consider the dynamic planning problem described above with the state space defined as $S = [0,1]^3$, where each state $s_t = (\theta_t, \tilde{L}_t, \omega_t)$, and the compact transformation $\tilde{L}_t = \frac{L_t}{1+L_t}$ ensures compactness. Let the compact, convex control set be $A = [0,1]^2$, with generic control variables $a_t = (p_t, m_t)$. Assume the instantaneous utility function
\begin{align*}
u(s,a) 
= \; & \lambda I_{\text{total}}(s,a) 
      - \gamma Var(I_{\text{total}}(s,a)) \\
    & - C_{\text{central}}(a)
      - \psi L
\end{align*}
is continuous and bounded, and the state transition function
\begin{align*}
s_{t+1}=f(s_t,a_t,\varepsilon_t,\nu_t),\quad\varepsilon_t\sim N(0,\sigma_{\varepsilon}^2),\quad\nu_t\sim N(0,\sigma_{\nu}^2)
\end{align*}
is measurable and continuous in $(s,a)$ almost surely. Further, denote the local governments' optimal response by $c^*(p,m)$, as derived explicitly in Lemma 1, which is continuous in the controls. Then, the following results hold:
There exists a unique bounded and continuous value function $V^*: S \to {R}$ satisfying the Bellman equation
\begin{align*}
V^*(s) = \max_{a\in A}{E}[u(s,a)+\delta V^*(f(s,a,\varepsilon,\nu))].
\end{align*}

There exist measurable policy functions $p^*(s), m^*(s): S\to[0,1]$ solving the central government's optimization problem.

The policy functions $(p^*(s), m^*(s))$, the local governments' optimal response $c^*(s)$, and the belief updating process $\theta_{t+1}(s,a)$ together constitute a Markov Perfect Equilibrium. Specifically, equilibrium conditions are satisfied such that:
\begin{itemize}
\item The central government's strategies solve the above Bellman equation given local governments' responses;
\item Local governments' responses are optimal given the central government's policies;
\item Market beliefs evolve consistently as per Bayesian updating mechanisms previously defined.
\end{itemize}
If the correspondence $A(s) = \arg\max_{a \in A} {E}[u(s,a) + \delta V^*(f(s,a,\varepsilon,\nu))]$ is upper hemicontinuous and has nonempty, convex values, then by the Kuratowski–Ryll-Nardzewski measurable selection theorem, there exists a Borel-measurable selector $a^*(s) = (p^*(s), m^*(s))$ satisfying
\begin{align*}
a^*(s) \in A(s) \quad \text{for all } s \in S.
\end{align*}
Moreover, under FOC and Slater-type interiority conditions on $u(s,a)$ and $C_{central}(a)$, the optimal policy satisfies the Karush-Kuhn-Tucker (KKT) conditions almost everywhere. A formal proof of existence, uniqueness, and measurability conditions is presented in Appendix.A. \\

Based on the result of Proposition 1, we proceed to examine the model's equilibrium properties.\\

\textbf{Corollary 1:} Under the conditions of Proposition 1 (such as discount factor $\delta \in (0,1)$, finite variance of shock terms, etc.), the optimal policy exhibits continuous and smooth properties:

\begin{itemize}
    \item If the central government persistently maintains relatively high narrative precision $p_t$ and monitoring intensity $m_t$, the local government's optimal ideal execution decision $c_t^*$ will drive market participants' belief $\theta_t$, and the expected belief gradually converges toward a steady-state credibility level $\theta^*$.
    
    \item Simultaneously, due to the increase in policy execution $c_t^*$, the proportion of skeptical market participants (Skeptics) $\omega_t$ will gradually decrease, eventually vanishing in the long run. Therefore, by appropriately designing $p_t$ and $m_t$, the central government can effectively promote credible institutionalization, enabling the market to transition from the Construct phase to the steady-state Commitment phase.
\end{itemize}

We further analyze, from the central government's perspective, how the heterogeneity in local governments' execution costs (i.e., variation in $k_{c,t}$) affects their optimal strategy. From Lemma 1, it is evident that the local government's ideal execution strategy $c_t^*$ responds positively to both the central government's monitoring intensity and narrative precision:
\begin{align*}
\frac{\partial c_{t,\text{ideal}}^*}{\partial p_t} > 0, \quad \frac{\partial c_{t,\text{ideal}}^*}{\partial m_t} > 0.
\end{align*}

Given the central government's period utility function:
\begin{align*}
U_S\bigl(S_t;\,p_t,m_t\bigr)
  &= \lambda\,\mathrm{E}\bigl[I_{\text{total},t}\bigr]
     - \gamma\,\mathrm{Var}\bigl(I_{\text{total},t}\bigr) \\
  &\quad - C_{\text{central}}\bigl(p_t,\,m_t\bigr)
     - \psi\,L_t
\end{align*}
where $S_t$ denotes the state variables. Since increasing monitoring intensity $m_t$ and narrative precision $p_t$ incurs marginal costs, the central government's strategy must balance the marginal cost with the marginal benefit derived from expected market investment returns and local governments' actual implementation outcomes.\\

\textbf{Proposition 2:} Given the distribution of local governments' heterogeneous capabilities $k_{c,t}$ and monitoring cost function $C_{\text{central}}(p, m)$, the central government's optimal strategy $(p_t^*, m_t^*)$ satisfies the following first-order conditions:

\begin{align*}
\frac{\partial U_S}{\partial p_t}
+ \delta\,\mathrm{E}\left[
  \frac{\partial V(S_{t+1})}{\partial p_t}
\right] &= 0, \\
\frac{\partial U_S}{\partial m_t}
+ \delta\,\mathrm{E}\left[
  \frac{\partial V(S_{t+1})}{\partial m_t}
\right] &= 0.
\end{align*}

Specifically, the central government adjusts its narrative precision and monitoring intensity to ensure that the marginal costs balance with the long-term marginal benefit of credibility improvement, thus forming a dynamically optimal strategy.

\subsection{Implication}
\pdfbookmark[2]{Implication}{sec:model}

the above model framework corresponds to the three stages of Narrative–Construct–Commitment framework. Each stage maps distinct dynamics:

In the Narrative stage, the central government sets $p_t$, i.e., Narratives, as the signaling device for policy objectives. This influences market expectations and directly determines the investment level of Believers, while also exerting pressure on local governments' policy execution. Local governments, in turn, choose execution consistency $c_t$ based on $p_t$ and monitoring intensity $m_t$, balancing implementation costs and administrative capacity. Over time, as credible Narratives are sustained, market participants' belief $\theta_t$ gradually increases, and Rational Participants' proportion rises while Skeptics' proportion shrinks. If such a process continues, market participants' beliefs eventually stabilize at a high level ($\theta^*$).

Simultaneously, higher execution consistency $c_t^*$ reduces market skepticism. As the proportion of Skeptics diminishes ($\omega_t \rightarrow 0$), the positive feedback loop between credible Narratives and effective execution gradually transitions the market from the Construct phase to the Commitment phase. The institutionalization process, represented by $L_t$, captures this transitionNarratives evolving into institutional commitment over time.

As $L_t$ accumulates, the improvement of execution consistency $c_t$ may further reduce the belief updating noise $\varepsilon_t$, bringing about sustained increases in $\theta_t$. Once institutional credibility stabilizes, local governments' execution stabilizes at a minimum guaranteed level $c_{\min}$, and deviations shrink further. Market skepticism vanishes in the long run, resulting in higher average investment and reduced economic volatility. Achieving this outcome requires the central government to maintain precise narrative signaling and strong monitoring, fostering market confidence while supporting credible institutional commitments. This embodies the Commitment phase's characteristics. Echoing Spence's (1973) Signaling Theory, capable administrators can credibly distinguish themselves through continuous policy signaling and effective execution, resulting in a Separating Equilibrium and yielding high long-term returns. The NCC framework thus establishes a self-enforcing dynamic mechanism where credible Narratives generate stable market expectations, reduced volatility, and persistent credibility updates. Hence, the three stages of NCC constitute a dynamic closed-loop mechanism of policy implementation and market belief updating.

\section{Empirical Strategy}
\pdfbookmark[1]{Empirical Strategy}{sec:empiracal}

In this section we outline our empirical work, namely how we evaluate our theoretical model using a local projection (LP) framework based on high-frequency identification.

\subsection{Why the Innovation-Driven Development Strategy as a Case}
\pdfbookmark[2]{Why the Innovation-Driven Development Strategy as a Case}{sec:empiracal}

We select the officially implemented \textit{Innovation-Driven Development Strategy} (IDDS) starting in 2016 as our case for empirical analysis and utilize data spanning from 2016 to 2023 (see IV for justification).

Firstly, from the Narrative perspective, IDDS represents a clear window of opportunity during the initial stage of policy signaling, where explicit dissemination of specific policy directives is observed. Since the 18th CPC National Congress proposed the "implementation of innovation-driven development strategy", the Central Committee and the State Council have continued to strengthen this narrative. For example, General Secretary Xi Jinping emphasized during the inspect in Harbin in May 2016, ``In order to implement the innovation-driven development strategy, we must focus on building a technological innovation system that is enterprise-based, market-oriented, and combines production, education, and research". On May 19, 2016, 18:56, the \textit{Outline of the National Innovation-Driven Development Strategy} was officially released, which clearly proposed that by 2020, an institutional environment and policy system that meets the requirements of innovation-driven development will be basically formed \footnote{Full announcement avaliable at: \url{https://www.gov.cn/zhengce/2016-05/19/content_5074812.htm}}. The clear definition and specific release date (i.e., precise time window) provide us with a clear basis for identifying time nodes and policy shocks when using high-frequency financial market data for event study.

Secondly, from the Construct perspective, IDDS has a significant policy dissemination and market expectation anchoring mechanism. Specifically, the strategy puts forward a number of supporting reform measures, covering specific industrial policies such as intellectual property protection, technical standard formulation, and the clarification of the dominant position of enterprises, aiming to strengthen the motivation and ability of independent innovation of enterprises through market-oriented reform. In 2015, The State Council issued \textit{Several Opinions on Deepening Institutional Reform and Accelerating the Implementation of Innovation-driven Development Strategy}, which especially emphasized the need to break market monopoly, implement strict intellectual property protection system and build a unified and transparent industrial technology standard system. These measures themselves are not specific commitments that take effect immediately, but construct a policy expectation that guides enterprises, capital and human resources to gradually transfer to the field of innovation. This process is consistent with the process in which the central government shapes market expectations through narrative precision and gradually guides resource allocation detailed in our theoretical model.

More critically, IDDS possesses the prototypical characteristics of Commitment. Unlike short-term policy shocks with temporary incentive effects, IDDS involves large-scale institutional arrangements and sustained input, with long-term objectives extending from 2020 to even longer horizons such as 2035 or 2050, aiming to establish a systematic institutional framework for innovation-driven development. The central government has set a target for stable R\&D investment (the proportion of R\&D investment in GDP will reach 2.55\% in 2022, maintaining sustained stability and increase thereafter), strengthened investment in science and technology funds and management reform, and ensured long-term stable resource allocation. Local governments have also introduced supporting measures to establish a long-term incentive mechanism for the transformation of scientific and technological achievements, an incentive mechanism for enterprise technological innovation, and a talent policy system. Such long-term policy commitment exactly reflects the dynamic mechanism of gradual accumulation and institutionalization of policy credibility in our theoretical model, which is reflected in the continuous positive promotion of economic structural transformation and total factor productivity (TFP) growth, which is reflected in the continuous and significant improvement of GDP, the proportion of innovation investment and technological progress indicators in macroeconomic data.

To analyze narrative shocks, the ideal policy case must satisfy two basic conditions: First, the policy objective should be clearly narrative and characteristic-explicit; second, policy announcements must exhibit clear exogeneity, facilitating the precise identification of the timing of policy shocks through event-study methods and high-frequency identification methods (Nakamura \& Steinsson 2018; Romer \& Romer 2004). The introduction of the IDDS provides us with such a theoretically ideal narrative shock identification scenario. By accurately identifying the timing of policy shocks using financial market instantaneous reactions, we can precisely determine the size and scope of narrative shocks. Based on the NCC framework, the narrative shock not only reflect the market's initial response to the initial narrative but also allows us to trace how market participants' expectations adjust over time.

Specifically, we regard the announcement on May 19, 2016 as the precision variable shock (\(p_t\)) of narratives in the theoretical model, and the immediate financial market responses at the time of announcements serve as the high-frequency measure of such narrative shocks(\(\nu_t\)). One of the core mechanisms in our model is the dynamic updating of market participants' evaluations of government credibility ($\theta_t$) during the policy implementation process. Although $\theta_t$ cannot be directly observed, we capture the real-time adjustment process of market expectations using financial market instantaneous responses (such as changes in stock indices within the window of policy announcements). This approach effectively maps the feedback loop of the dynamic Bayesian updating mechanism for market participants (rational participants and skeptics) in the model. Finally,  when credibility accumulates to a threshold value, policies will achieve institutional commitment ($L_t$), entering a phase of sustained stable implementation. We have also established corresponding observation criteria in our empirical analysis, observing whether policy narrative shocks can, after generating immediate responses to market participants' expectations (Construct phase), further continuously impact real economic indicators (such as innovation activities, investment, production, etc.) to form long-term stable trend changes. Through the impulse response functions (IRFs) of the empirical model, we can directly test the existence of this long-term commitment effect.

\subsection{Narrative Shocks Identification}
\pdfbookmark[2]{Narrative Shocks Identification}{sec:empiracal}
Our identification strategy draws upon the high-frequency identification (HFI) approach proposed by Nakamura \& Steinsson (2018), among others. The fundamental logic is that when an important macro policy is announced within an extremely short time window, the immediate impact on market prices can reasonably be regarded as primarily driven by the policy information itself, rather than being significantly affected by other macroeconomic shocks occurring simultaneously (Nakamura \& Steinsson 2018; Gürkaynak, Sack \& Swanson 2005). Therefore, if we observe significant price jumps within the according time window the policy announcement, we can interpret them as the direct impact of policy narrative information (Narrative) on market expectations, and subsequently define them in regressions as exogenous narrative shocks for causal identification purposes.

To ensure the rigor of this identification logic, we rely on the following two key assumptions.

Firstly, the Narrow-Window Exogeneity Assumption. That is, within the extremely short time window surrounding the policy announcement (e.g., ±30 minutes), the movements in market asset prices can reasonably be viewed as driven solely by the new information conveyed by the announcement, rather than by other economic or financial factors. Specifically, most macroeconomic data releases (such as GDP, CPI, industrial value-added) have pre-fixed release dates and are fully anticipated by the market, which implies that these economic indicators themselves would not provide additional new information shocks to the market within the extremely short time window of a policy announcement. Even if the market's interpretation of certain macro data is not fully complete, it usually requires several hours or even days of discussion, digestion, and repricing processes. Therefore, asset price volatility observed within such a narrow window can fully reflect investors' immediate digestion and assessment of the new information contained in the policy announcement (romer \& romer 2010; Nakamura \& Steinsson 2018).

Secondly the Unanticipated Content Assumption. This assumption does not imply that the policy itself is completely unexpected by the market (in fact, the market often anticipates the timing and general direction of the policy); rather, it emphasizes that the market cannot precisely predict the specific wording, implementation intensity, and concrete fiscal or institutional arrangements contained in the announcement. It is precisely these detailed elements that determine the actual impact of the policy announcement on market expectations: if all content had been fully anticipated, market prices would not exhibit significant jumps during the announcement window. Therefore, the market faces unanticipated content shocks at the moment of the announcement, prompting a rapid adjustment of expectations regarding future policy credibility and economic growth potential, resulting in the observed instantaneous price movements (Romer \& Romer 2004).

In the context of our study, although the IDDS was first conceived during the 18th CPC National Congress in 2012 and has undergone considerable groundwork, we argue that the formal document issued by the State Council in 2016 still contains a large amount of specific, operational arrangements. These key details could not have been precisely anticipated by the market prior to the policy release. Thus, the immediate market reactions we observe directly reflect the market's rapid reassessment of policy credibility, implementation intensity, and future economic growth potential.

One technical challenge for high-frequency identification studies is that they typically assume policy announcements occur during financial market trading hours, whereas in China, policy announcements are often released after market close. To address this, we follow the Opening Gap strategy proposed by Bernanke \& Kuttner (2005). Specifically, we define the Narrative Shock as the difference between the minute-level moving average of market prices during the last 30 minutes before market close on the last trading day prior to the announcement, and the minute-level moving average of market prices during the first 30 minutes after market opening on the next trading day following the announcement:
\begin{align*}
\nu_{t^*}^{\text{Narrative}} = \frac{P_{t^*+1}^{\text{open}, \, MA(k)} - P_{t^*}^{\text{close}, \, MA(k)}}{P_{t^*}^{\text{close}, \, MA(k)}},
\end{align*}
where $t^*$ denotes the last trading day before the policy announcement (i.e., May 19), and $t^*+1$ denotes the first trading day after the policy announcement (i.e., May 20). $P_{\text{close}}$ and $P_{\text{open}}$ refer to the moving average prices within the $k$-minute window before market close and after market open, respectively. We adopt $MA(10)$ to balance the market's information response speed and price smoothness. We choose this method not only to avoid the issue of relying solely on the difference between opening and closing prices but also because the moving average method effectively smooths short-term noise and liquidity shocks, ensuring the measurement accuracy of policy narrative shocks and information effect. At the same time, this definition strictly follows the exogeneity requirement of high-frequency identification, considering that the T+1 regulation and the disclosure of policy announcements occurs outside of trading hours. Market investors are unable to trade at the moment the policy announcement is released. The price variation within the short window after market open the next day truly reflects the first-time valuation adjustment of market participants in response to the narrative rather than interference from other information. Since the market has not yet obtained specific policy information at the close of the previous day, such method can clearly measure the instantaneous causal effect.

The above discussion clarifies the definition and causal inference framework of our high-frequency narrative shock strategy. It will serve as the core exogenous explanatory variable to portray how the narrative (IDDS) affects market participants' expectations and real economic outcomes. We will provide further justification on the construction of the Narrative Shock variable in IV.A.

\subsection{Impulse Responses by Local Projections}
\pdfbookmark[2]{Impulse Responses by Local Projections}{sec:empiracal}

In order to capture the dynamic transmission mechanism of the Narrative Shocks, specifically how narrative information affects the medium to long-term decisions of economic agents and ultimately leads to changes in real economic outcomes, we employ the Local Projection (LP) method to depict the dynamic response path of policy narrative shocks to output variables. Specifically, we construct the following LP regression framework based on the classic model by Jordà (2005) and Stock \& Watson (2018):
\begin{align*}
Y_{t+h} - Y_t =\; 
& \alpha_h 
+ \beta_h \nu_t^{\text{Narrative}} 
+ \Gamma_h X_{t-1} \\
& + \gamma_h \cdot \text{trend}_t 
+ \delta_{\text{COVID},h} D_{\text{COVID},t} 
+ \varepsilon_{t+h}
\end{align*}
where $Y_{t+h}$ represents the output level $h$ periods after the policy shock occurs. $\nu_t^{\text{Narrative}}$ denotes the exogenous Narrative Shock identified using our high-frequency identification, satisfying the exogeneity and unanticipated content assumption. The control variable vector $X_{t-1}$ includes possible macroeconomic covariates lagged by one period to capture the pre-policy dynamics. $\Gamma_h$ represents the coefficient vector for control variables, used to mitigate omitted variable bias. In practice, we control for pre-existing trends or fluctuations (sample period fixed effects: 1, 2, 3, …). $D_{\text{COVID}}$ is the dummy variable for pandemic shocks, with coefficient $\delta_{\text{COVID},h}$ capturing the impact of the pandemic. $\gamma_h$ is the coefficient for the linear trend term. $\beta_h$ is the coefficient of interest, measuring the dynamic response of macro variables to the narrative shock. $\varepsilon_{t+h}$ is the error term satisfying standard conditions. Detailed variable definitions are provided in IV.B. This section focuses on the overall estimation framework of the Local Projection method.

We adopt the Local Projection method based on the following considerations. Firstly, compared to the traditional structural vector autoregression (SVAR) models, LP does not require imposing contemporaneous identifying restrictions on the covariance structure of endogenous variables or restrictions on the Data Generating Process (DGP), thereby avoiding the risk of potential model misspecification stemming from strict and potentially unrealistic structural assumptions often necessary in SVAR (Ramey 2016; Stock and Watson 2018). This is particularly important for fiscal and industrial policies analysis, we cannot reasonably assume that innovation activities have an immediate, one-way causal effect on total factor productivity (TFP) or investment, nor can we rule out the possibility that GDP changes in turn immediately affect innovation and investment decisions. These problems cannot be effectively avoided in the SVAR's underlying contemporaneous structure (Plagborg-Møller \& Wolf 2021). Secondly, Local Projection naturally accomodates exogenous shocks obtained by high-frequency identification. We can directly include it in the linear regression as an independent exogenous variable without the need to use a complex structural identification process like SVAR, which significantly improves the transparency and robustness of identification (Jordà 2005; Stock and Watson 2018). Thirdly, under the LP framework, the shock response coefficient $\beta_h$ for each future period h is estimated separately. This period-by-period regression method has a more controllable estimation variance and can flexibly detect and handle the nonlinearity and state dependence nature of the shock, and potential observer-expectancy effect (Auerbach \& Gorodnichenko 2012). Meanwhile, LP can obtain robust IRFs as long as Narrative Shock satisfies the exogeneity condition described above (Nakamura \& Steinsson 2018; Gürkaynak et al. 2005).

As mentioned earlier, though narrative shocks are strictly defined as immediate market reactions within the policy announcement window, this by no means indicate we ignore potential pre-announcement effects. In fact, since IDDS was first proposed in 2012, followed by the issuance of various opinions in 2015, and the formal introduction of the outline in 2016, there has indeed been a long-term foundation for policy narratives, and market expectations may have partially formed (Ramey 2011). However, the formal introduction of the outline still provided significant incremental information regarding policy details, implementation pathways, and government commitment credibility, especially at the timing of specific rules and authoritative announcements, taking a critical step from relatively ambiguous narrative to institutional commitment, which we particularly reclarify here.

Finally, based on the aforementioned LP framework, by estimating $\beta_h$ individually, we obtain the Impulse Response Function (IRF) of the narrative shock generated by the release of the Innovation-Driven Development Strategy outline at each future time point, constituting the complete dynamic impact pathway of Narrative Shock on macroeconomic variables:
\begin{align*}
IRF(h) = \frac{\partial y_{t+h}}{\partial \nu_{t}^{\text{Narrative}}} = \beta_h,
\end{align*}

Under the conditions of strict exogeneity requirement ${E}[\varepsilon_{t+h}|\nu_t^{Narrative},X_{t-1}] = 0$ and weak stationarity, the $\hat{\beta}_h$ obtained through OLS estimation is a consistent estimator of $\beta_h$, meaning that the dynamic response path estimated by the LP method converges with the true impulse response path in large samples (Plagborg-Møller \& Wolf 2021). In the estimation, to robustly handle potential serial correlation and heteroscedasticity, we will adopt the Newey-West heteroscedasticity and autocorrelation consistent (HAC) standard error estimation method (Newey \& West 1987) to ensure the validity and robustness of statistical inference.

\section{Estimation and Results}
\pdfbookmark[1]{Estimation and Results}{sec:Estimation}

In this section we discuss our dataset, LP estimations, and results specifically.

\subsection{Narrative Shock}
\pdfbookmark[2]{Narrative Shock}{sec:Estimation}

In our theoretical model in II, we assumed heterogeneous market participants divided according to belief processing differences (Believers, Rational Participants, Skeptics), and incorporated risk-neutral settings, thus allowing different participants to have divergent speeds and directions in response to policy shocks. However, heterogeneity in reality is often not only vertical (based on cognitive or behavioral differences) but also includes horizontal dimensions, meaning that different sectors and types of listed companies do not respond with the same intensity or manner to policy information. Therefore, we construct high-frequency narrative shocks using the following financial market variables classified by enterprise size and industry characteristics:
\begin{itemize}
    \item CSI300 Index. As one of the most representative large-cap indices in China's stock market, the CSI300 can effectively capture the market's overall immediate expectation updates and risk preference changes in response to macroeconomic policy shocks, representing the macroeconomic expectations of the entire market;
    \item 50ETF Index. This index fund tracks the constituent stocks of the SSE 50 Index, mainly covering large market capitalization blue-chip enterprises, and can more precisely reflect the immediate reactions of institutional investors and large market participants, tending toward a rational or medium to long-term pricing perspective;
    \item ChiNext Index. Since the innovation-driven strategy is expected to have a greater impact on high-tech enterprises and growth enterprises, we specifically include the ChiNext Index to capture immediate expectation changes regarding the market's long-term innovation investment and technological innovation incentive policies.
\end{itemize}

Risk neutrality implies that no single entity has an overweighted position in price information, nor is there a prior determination of which sector best represents the ``true shock." Therefore, without introducing additional complex weighting assumptions or component analysis, we use equal-weight processing:
\begin{align*}
\nu_{t}^{\text{Narrative}} = \frac{1}{3} \left( \nu_{\text{CSI300},t} + \nu_{\text{50ETF},t} + \nu_{\text{ChiNext},t} \right).
\end{align*}

We view each sector as ``equally important" or having ``equal pricing rights," without favoring any single market domain. Using market capitalization weighting or liquidity weighting would introduce another dimension of assumptions (such as ``market importance" with market value as a proxy), which has no direct mapping relationship with our theoretical model, and would instead make the results susceptible to subjective weight fluctuations. Equal weighting performs dimensionality reduction and preprocessing of potential noise variables at the shock measurement level, accommodating the model's brevity,  robustness of causal inference, and clarity of statistical significance. It should be clarified that ``shock" refers to the absolute intensity of market fluctuations after policy release, representing the scale of uncertainty or cognitive shock, rather than specific favorable/unfavorable directions.

According to calculation results, the shock magnitudes of the three indices are: 0.1721\% (CSI300 Index), 0.2463\% (ChiNext Index), and 0.0435\% (50ETF Index). Subsequently, we simply and effectively calculated the equally weighted average of these three indices' shock magnitudes to obtain the final Narrative Shock Index of 0.154\%.

We further introduce a Strengthened Narrative setting to simulate the continuous transmission and periodic reinforcement of IDDS narratives at specific key points in time. Specifically, in 2017 and October 2022, corresponding to the 19th and 20th CPC National Congresses, considering the reinforcing effect of major political events on market expections, we introduced additional enhanced shocks of 0.0001 in those quarters, which is still based on considerations of examining the immediate, short-term, medium-term, and even long-term dynamic impacts (impulse response) of the shock. It is worth emphasizing that although narrative shocks formally appear as a quarterly dummy-type variable, they actually represent structural adjustments of beliefs and expectations caused by specific policy information.

\subsection{Local Projection Redux}
\pdfbookmark[2]{Local Projection Redux}{sec:Estimation}

Now we explain our LP specification in full detail.
\begin{align*}
Y_{t+h} - Y_t =\; 
& \alpha_h 
+ \beta_h \nu_t^{\text{Narrative}} 
+ \Gamma_h X_{t-1} \\
& + \gamma_h \cdot \text{trend}_t 
+ \delta_{\text{COVID},h} D_{\text{COVID},t} 
+ \varepsilon_{t+h}
\end{align*}
\textbf{Dependent Variables $(Y_t)$.} We selected six core economic variables to examine the dynamic effects, namely quarterly real GDP, labor productivity growth rate, fiscal technology expenditure, manufacturing fixed asset investment growth rate, government consumption expenditure as a percentage of GDP, and industrial value added as a percentage of GDP. These variables were chosen because they clearly and accurately correspond to different stages in our theoretical NCC framework model:

\begin{itemize}
\item GDP, Labor Productivity Growth: Reflect the final growth effects and efficiency improvements in the theoretical model, intuitively revealing the overall economic impact of policies, especially demonstrating the effects of economic structural transformation and total factor productivity (TFP) growth after policy implementation in the Commitment stage;
\item Fiscal Technology Expenditure: Captures the direct short-term input changes of the government public sector after policy narrative adjustments in the Construct stage, reflecting whether the government clearly increases resource allocation to the technology sector in the short term;
\item Manufacturing Fixed Asset Investment Growth: More specifically characterizes changes in enterprise-level investment behavior, representing the specific actions of enterprise participants after policy narratives and belief changes, serving as an important connecting indicator between the Construct and Commitment stages of policy;
\item Government Final Consumption as a Percentage of GDP, Industrial Value Added as a Percentage of GDP: Reflect the long-term manifestation of economic structural changes after policy implementation, demonstrating whether industrial policies can significantly guide the direction of economic resource allocation in the medium to long term (structural transformation from Construct to Commitment).
\end{itemize}

\textbf{Core Explanatory Variable $(V_t^{Narrative})$.} Obtained using high frequency identification introduced in IV.A.

\textbf{Control Variables $(X_{t-1})$.} We rigorously selected the following important macroeconomic control variables, included in the regression with a one-period lag to effectively avoid the omitted variable bias (OVB) problem, and to control for exogenous shocks in the international economic environment:

\begin{itemize}
\item SHIBOR 3-month rate: Monetary policy environment indicator, representing overall financing costs and short-term monetary conditions;
\item M2 growth rate: Changes in broad money supply, reflecting monetary environment, credit expansion, and financial conditions;
\item Dollar Index (DXY) and RMB to USD exchange rate (USDCNY): Represent international macroeconomic environment and exchange rate fluctuation factors, effectively capturing the impact of external economic shocks and international economic environment fluctuations on domestic macroeconomics.
\end{itemize}

Additionally, we specifically included a linear time trend term (trend) and a dummy variable for the COVID pandemic period. The time trend term $trend_t$ is defined as a quarterly sequence (with the sample starting point 2016Q1 corresponding to a value of 1, and increasing by 1 for each subsequent quarter), to control for potential long-term trends in economic variables, non-stationarity, and slowly evolving structural changes. The pandemic period dummy variable $D_{COVID,t}$ is defined as having a value of 2 from the second quarter of 2020 to the fourth quarter of 2020, a value of 1 from the first quarter of 2021 to the fourth quarter of 2022, and 0 otherwise, specifically capturing abnormal fluctuations during the COVID-19 pandemic period.

It is also worth mentioning the introduction and implementation of the New Quality productive forces (NQPFs) initiative,which was first conceived in September 2023 and gradually formalized thereafter. Because there is a certain degree of policy overlap with IDDS, our data start from the first quarter of 2016 and ends in the fourth quarter of 2023 to avoid confusion in causal inference caused by overlapping effects. To conclude, in our local projections approach, the shock effect ($\beta_h$) for each future forecast period is associated with subsequent economic variables, thus enables us to identify the inter-temporal transmissive mechanism of high-frequency narrative shocks in low-frequency economic variables. 

\subsection{Analysis and Interpretation}
\pdfbookmark[2]{Analysis and Interpretation}{sec:Estimation}

\begin{table*}[t] \label{t1}
\small 
\centering
\caption{Significant Impulse Response Coefficients by Horizon}
\label{tab:irf_coefficients}
\setlength{\tabcolsep}{5pt} 
\renewcommand{\arraystretch}{1.2} 
\begin{tabular}{lcccccccccc}
\hline
\multicolumn{11}{c}{\textit{Part 1: Horizon 1--12}} \\
\hline
Horizon & 1 & 2 & 3 & 4 & 6 & 7 & 8 & 10 & 11 & 12 \\
\hline
GDP & -- & -- & -- & -- & -- & -- & \textbf{0.0248} & -- & -- & -- \\
LPG & -- & -- & -- & -- & -- & -- & -- & -- & -- & -- \\
ET  & -- & \textbf{0.7947} & -- & 1.0431 & \textbf{0.9758} & -- & 1.1981 & \textbf{1.0998} & -0.2587 & 1.0210 \\
FAI & \textbf{-0.0449} & -- & -- & -- & -- & -- & -- & -- & -- & -- \\
GC  & 0.0006 & \textbf{0.0007} & 0.0004 & -- & -- & -- & -- & -- & -- & -- \\
IVA & \textbf{-0.0030} & -0.0033 & -- & -- & -- & -0.0024 & -- & \textbf{0.0056} & -- & -- \\
\hline
\end{tabular}

\vspace{4mm}

\begin{tabular}{lcccccccccc}
\hline
\multicolumn{11}{c}{\textit{Part 2: Horizon 13--22}} \\
\hline
Horizon & 13 & 14 & 15 & 16 & 17 & 18 & 19 & 20 & 21 & 22 \\
\hline
GDP & \textbf{0.0838} & \textbf{0.0271} & \textbf{-0.0671} & \textbf{-0.0403} & \textbf{-0.0448} & \textbf{-0.0311} & \textbf{0.0588} & -- & -- & \textbf{0.0452} \\
LPG & \textbf{0.0240} & \textbf{0.0207} & \textbf{0.0138} & -- & -- & \textbf{-0.0287} & \textbf{-0.0171} & \textbf{-0.0056} & -- & \textbf{0.0223} \\
ET  & -- & \textbf{1.1088} & -0.2305 & \textbf{1.3773} & -- & \textbf{1.0661} & -0.2870 & \textbf{1.0982} & -- & -- \\
FAI & \textbf{0.1631} & \textbf{0.1147} & \textbf{-0.0998} & \textbf{-0.1052} & \textbf{-0.1489} & \textbf{-0.1555} & \textbf{0.0360} & -0.0057 & \textbf{-0.0278} & -- \\
GC  & -- & \textbf{-0.0004} & \textbf{-0.0003} & -- & \textbf{0.0009} & \textbf{0.0016} & \textbf{0.0010} & \textbf{0.0006} & \textbf{0.0004} & \textbf{0.0003} \\
IVA & \textbf{0.0040} & \textbf{0.0064} & \textbf{0.0043} & -- & -- & \textbf{-0.0031} & -- & -- & \textbf{-0.0029} & \textbf{-0.0064} \\
\hline
\end{tabular}

\vspace{2mm}
\begin{minipage}{0.95\textwidth}
\footnotesize
\textit{Note}: Bold entries indicate significance at the 1\% level. Other numbers denote either significance at the 10\% level or are not statistically significant. Abbreviations: LPG = Labour Productivity Growth; ET = Expenditure in Manufacturing; FAI = Fixed Asset Investment; GC = Government Final Consumption/GDP; IVA = Industry Value Added/GDP.
\end{minipage}
\end{table*}

Admittedly, monthly or even higher frequency data could identify the micro-timing and mechanisms of policy implementation and test the robustness of our model more precisely. However, constrained by existing macroeconomic data collection, most key economic variables (such as GDP, fiscal expenditure, labor productivity, and investment) can only achieve quarterly frequency in existing data. This limitation is not unique but a common issue in existing literature (Mertens \& Ravn 2013; Nakamura \& Steinsson 2018). Moreover, the limitation of time span is also one of the challenges faced. As a relatively new policy framework, IDDS has limited implementation time, resulting in constraints on the number of time series observations available for analysis. Although ideally, industrial policy research typically requires data spanning a longer time frame to fully capture its long-term effects, considering the dynamic nature of policy evolution and to avoid interference effects between different policy cycles, we are certain that the current time window setting and estimation results still pocess rationality and robustness in methodology and our model. We fully recognize the statistical inference challenges that sample size limitations may bring, therefore we have specifically adopted corresponding adjustment strategies in the model design, including using more streamlined parameterization methods and robust statistical testing procedures to minimize potential biases from small samples. We believe that with the passage of time and data accumulation, future research will be able to further test the rigor of the NCC framework and its robustness in the real world from an empirical perspective based on richer time series observations\footnote{Main results are presented for brevity consideration, full data provided in the the appendix.}. 

\begin{figure*}[t]
\centering
\begin{minipage}[b]{0.48\linewidth}
    \centering
    \includegraphics[width=\linewidth]{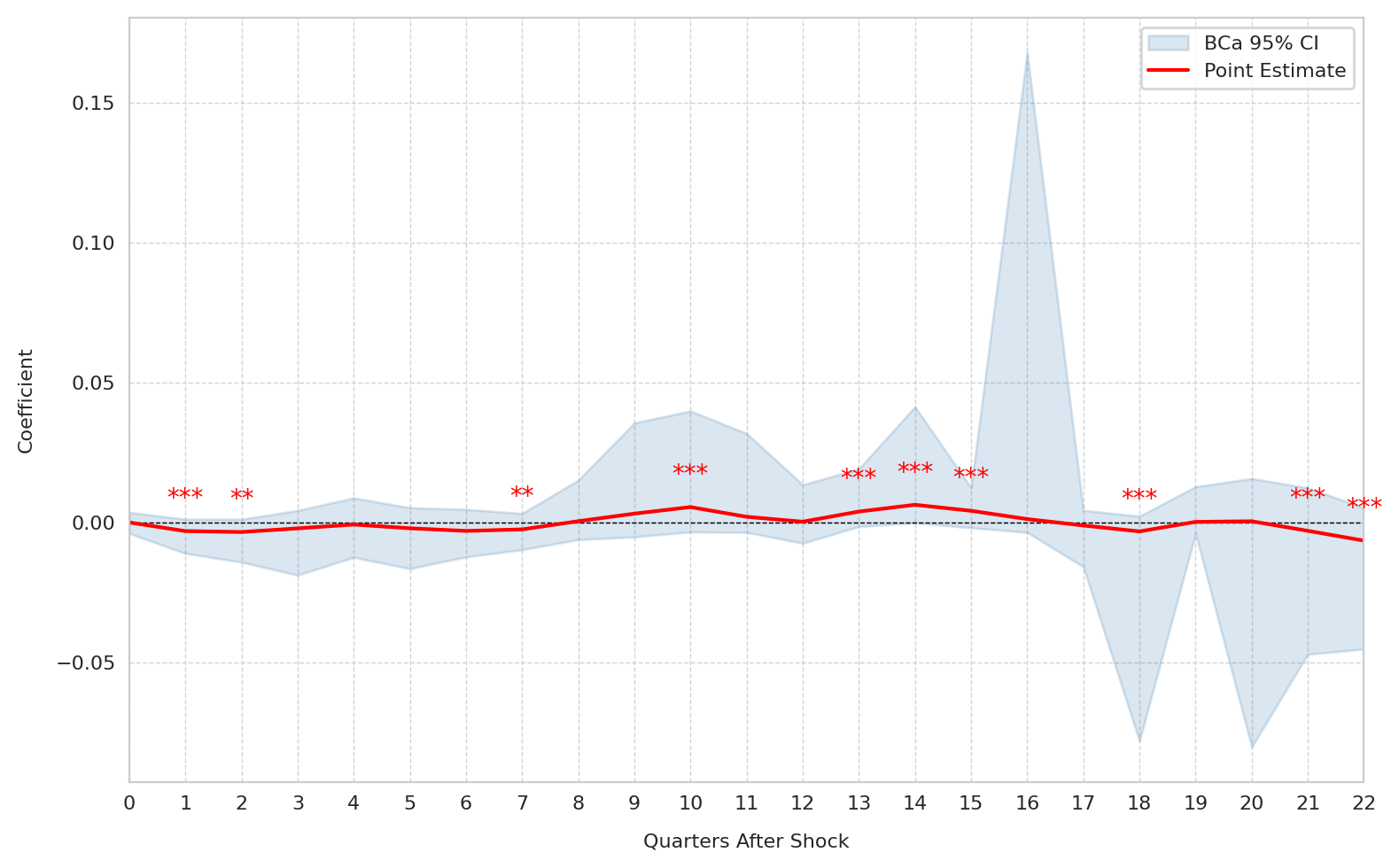}
    \caption{Impulse Response of GDP}
    \label{fig:gdp_response}
\end{minipage}
\hfill
\begin{minipage}[b]{0.48\linewidth}
    \centering
    \includegraphics[width=\linewidth]{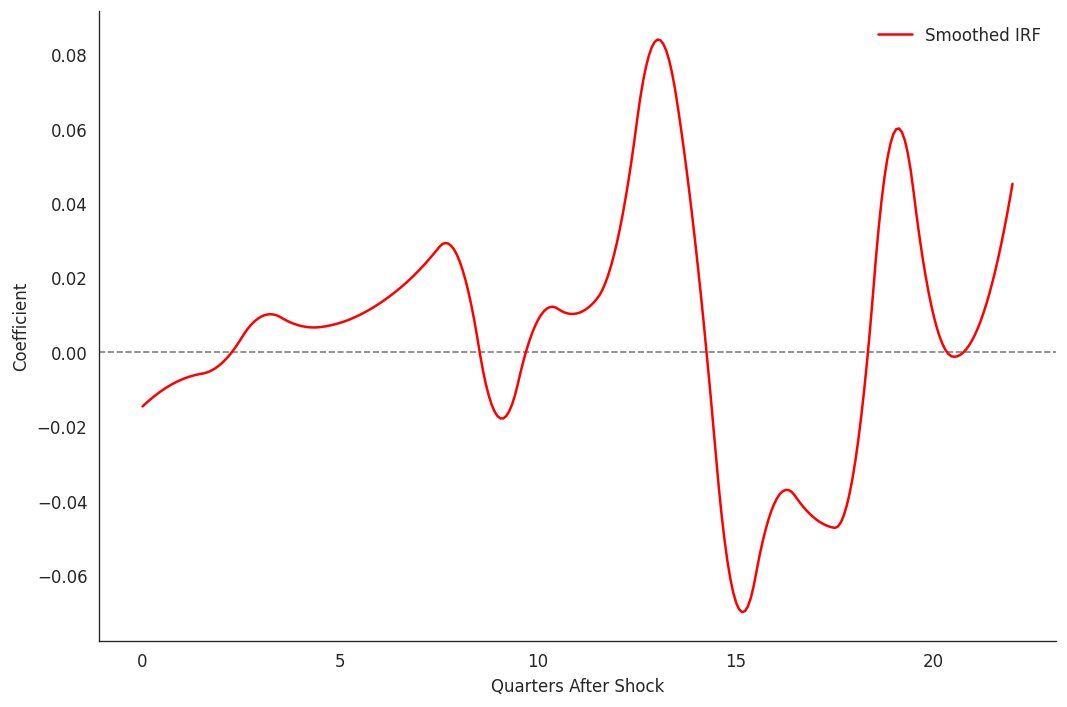}
    \caption{Smoothed Impulse Response of GDP}
    \label{fig:gdp_smooth}
\end{minipage}
\end{figure*}

\vspace{5mm}

\begin{figure*}[t]
\centering
\begin{minipage}[b]{0.48\linewidth}
    \centering
    \includegraphics[width=\linewidth]{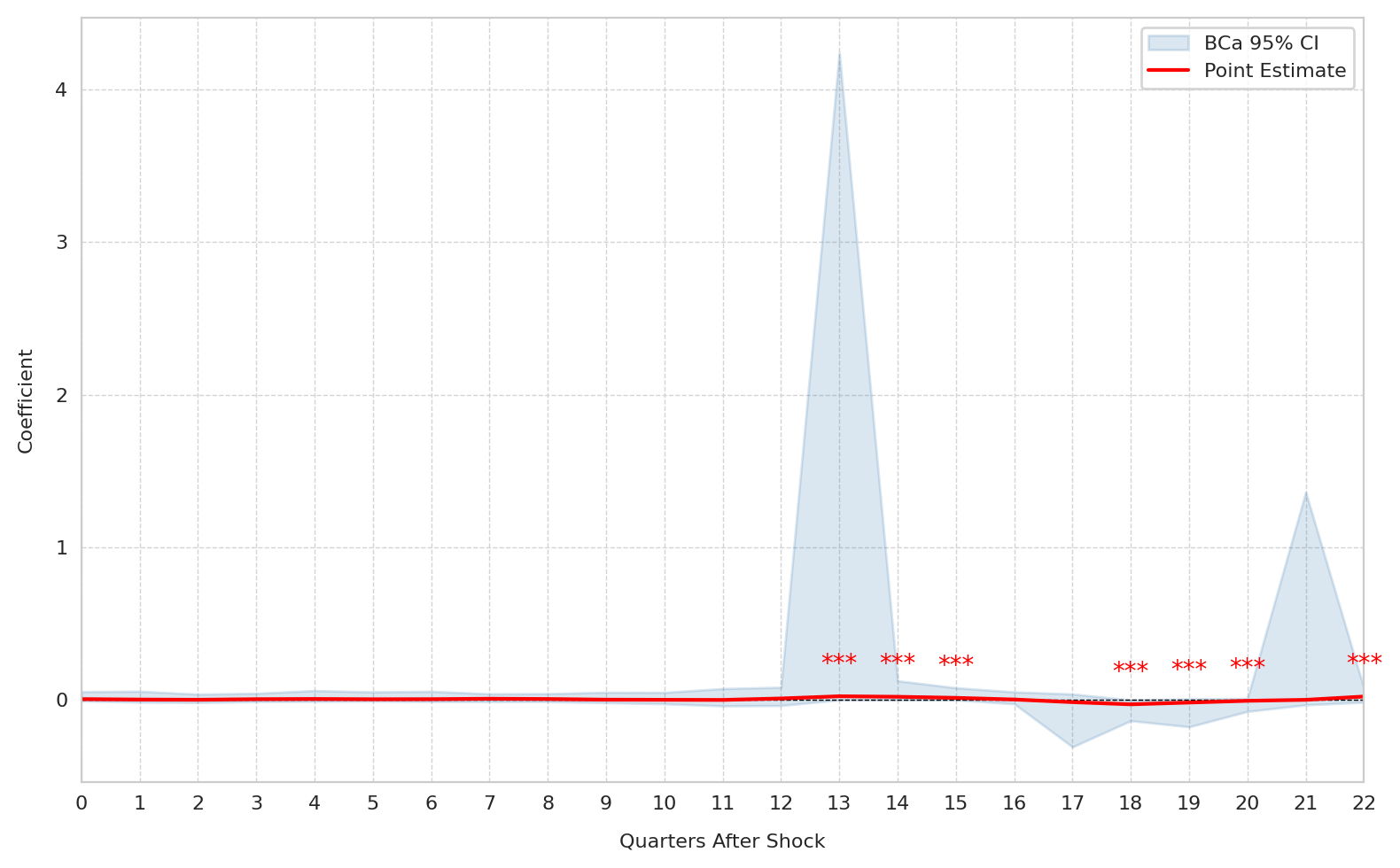}
    \caption{Impulse Response of Labour Productivity}
    \label{fig:lp_response}
\end{minipage}
\hfill
\begin{minipage}[b]{0.48\linewidth}
    \centering
    \includegraphics[width=\linewidth]{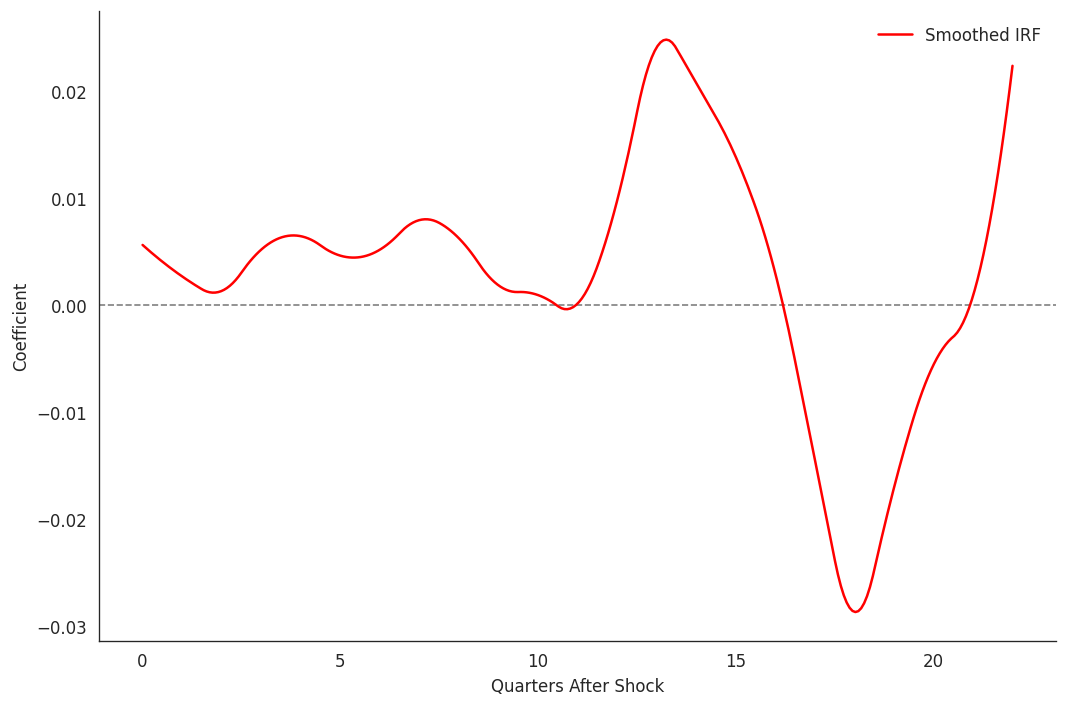}
    \caption{Smoothed Impulse Response of Labour Productivity}
    \label{fig:lp_smooth}
\end{minipage}
\end{figure*}

Table~\ref{t1} presents the estimated impulse responses across different forecast horizons for key macroeconomic indicators, including GDP growth, labor productivity growth, technology expenditures (ET), fixed asset investment (FAI), government consumption (GC), and industrial value added (IVA). First, consistent with the theoretical expectations from our NCC framework, GDP growth does not exhibit significant short-term responses. The earliest statistically significant positive response occurs at horizon 8, suggesting a lagged and gradual transmission mechanism from initial narrative shocks to observable economic outcomes. Such delayed effects are expected given the time required for narratives to transform social expectations into tangible economic activity, which is a typical characteristics of industrial policy and is consistent with the mechanism of the NCC framework. Notably, at longer horizons, GDP growth estimates exhibit volatility, with a pronounced negative response emerging around horizon 15. It is important to highlight that this significant negative response aligns precisely with the first quarter of 2020, coinciding with the onset and peak severity of the COVID-19 pandemic outbreak. Consequently, this specific downturn should likely be interpreted not purely as a result of the narrative shock, but rather as an exogenous disruption from the pandemic. Labor productivity growth (LPG) responses across most horizons are relatively muted, indicating that the primary transmission channel of policy narratives likely does not directly or immediately operate through productivity enhancements. Instead, narrative shocks might initially reallocate resources or shift investment patterns without immediately impacting productivity levels, which typically require longer structural adjustments. Technology expenditures (ET), reflecting government commitments to innovation and structural transformation, respond robustly at multiple horizons (particularly horizons 2, 4, 6, and 10), confirming the anticipated rapid and direct effect of policy narratives on government fiscal behavior. This pattern strongly supports the theoretical prediction that narratives effectively signal future credible government commitments, prompting quick adjustments in fiscal priorities and expenditures toward targeted policy goals. Fixed asset investment (FAI) displays a notable initial contraction followed by fluctuations at intermediate horizons. The sharp negative response at horizon 1 may reflect immediate market uncertainty following a narrative announcement. Subsequently, the volatility in FAI around horizons 15–18 corresponds is also consistent with the onset of the COVID-19 pandemic (Q1 2020 onwards). Responses of government consumption (GC) and industrial value-added (IVA) to narrative shocks appear comparatively small and sporadic in significance, suggesting these channels may not constitute the primary avenues through which narratives influence economic performance in the short-to-medium term.

\begin{figure*}
\centering
\begin{minipage}[b]{0.48\linewidth}
    \centering
    \includegraphics[width=\linewidth]{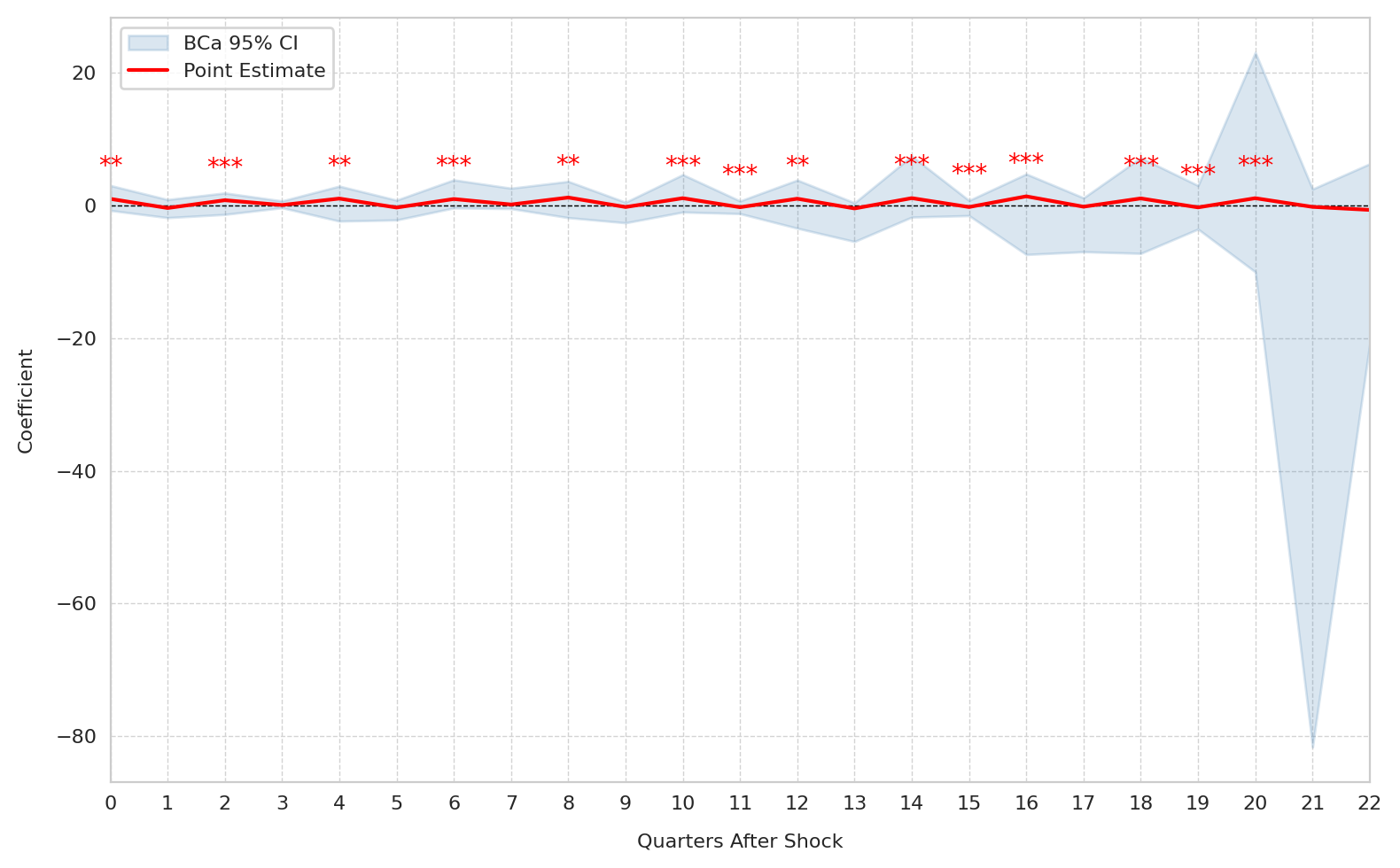}
    \caption{Impulse Response of Technology Expenditure}
    \label{fig:TE_response}
\end{minipage}
\hfill
\begin{minipage}[b]{0.48\linewidth}
    \centering
    \includegraphics[width=\linewidth]{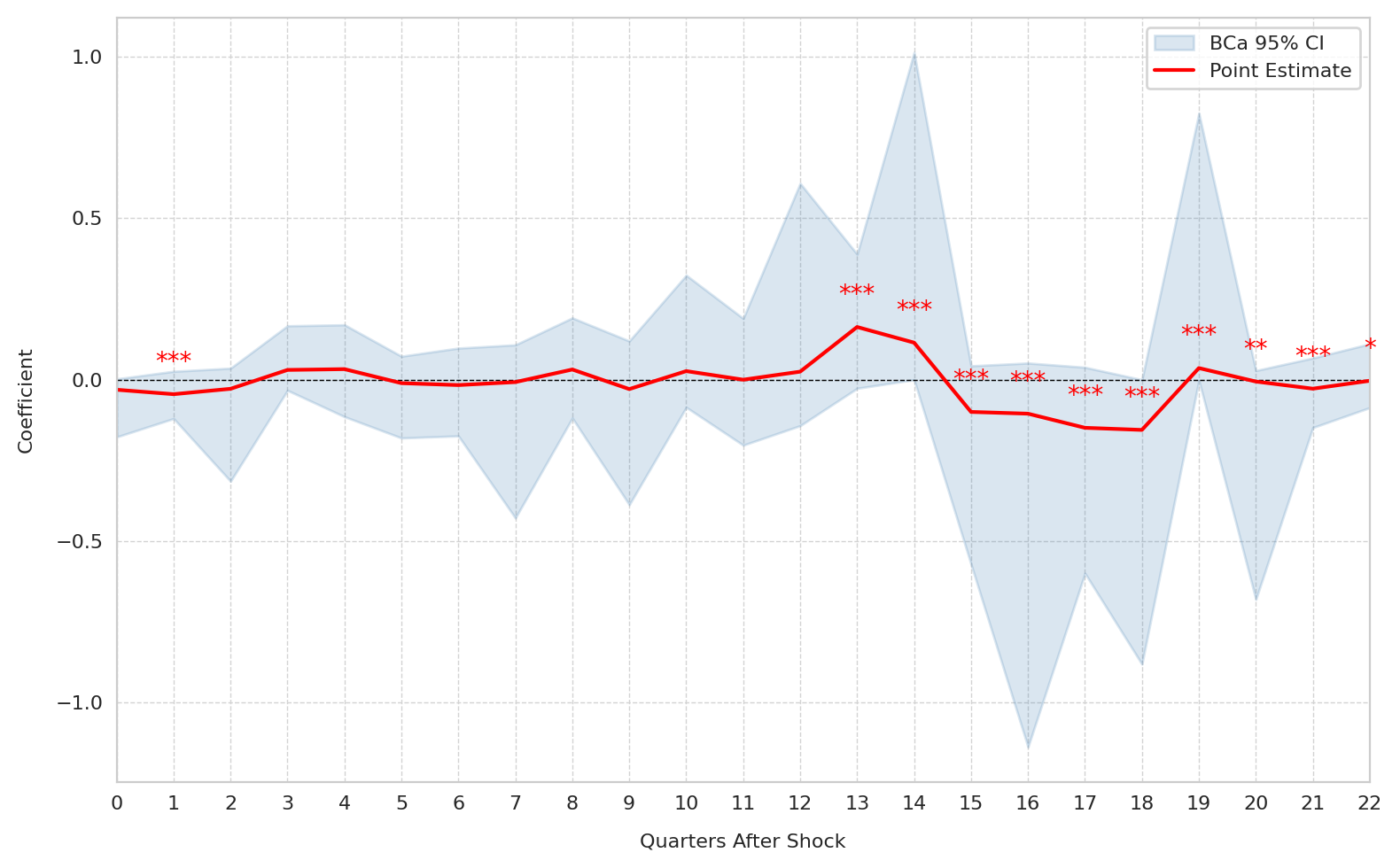}
    \caption{Impulse Response of Manufacturing Investment}
    \label{fig:FX_RESPONSE}
\end{minipage}
\end{figure*}

\textbf{GDP \& Labor Productivity.} Looking at the impulse response functions (IRFs) for GDP growth and labor productivity specifically (Figure 1 to 4). These two fundamental indicators directly linked to the core question of our study which is the impact of policy narratives on economic growth through their influence on productivity. One of the most striking features is the wide confidence bands at certain horizons, especially around quarters 15–21 for GDP and quarters 13 and 21 for labor productivity. These large standard errors merit particular attention. In our theoretical framework, the Construct-to-Commitment transition implies that agents only gradually update their beliefs about long-term policy credibility. Around the time those beliefs finally shift in a decisive way (i.e., once the policy narrative has built enough momentum to be taken as committed), we expect heightened uncertainty, as some investors or firms may remain skeptical, while others revise their expectations sharply upward. That discrepancy in how market participants react can generate volatile responses in the aggregate data, which, in turn, translates into larger standard errors in the IRFs.

Moreover, our theoretical model's Bayesian updating mechanism predicts that once new signals about policy credibility are strong enough to tip collective expectations, the posterior beliefs (and thus firms' and households' behaviors) can exhibit sudden jumps. If such jumps coincide with external shocks or discontinuous information events, one observes both large point estimates and broad confidence intervals. In Figure 1 (GDP), for instance, the confidence band spikes precisely when the economy might have been experiencing both narrative-induced expectation shifts and pandemic-related turmoil which result in uncertainty. Meanwhile, for productivity in Figure 3, wide bands around quarter 12–14 also reflect a confluence of factors: ongoing adjustments to the narrative shock (which is slower to impact productivity than GDP) and possible external disruptions that distort measured productivity in the short run. 

\begin{figure*}
\centering
\begin{minipage}[b]{0.48\linewidth}
    \centering
    \includegraphics[width=\linewidth]{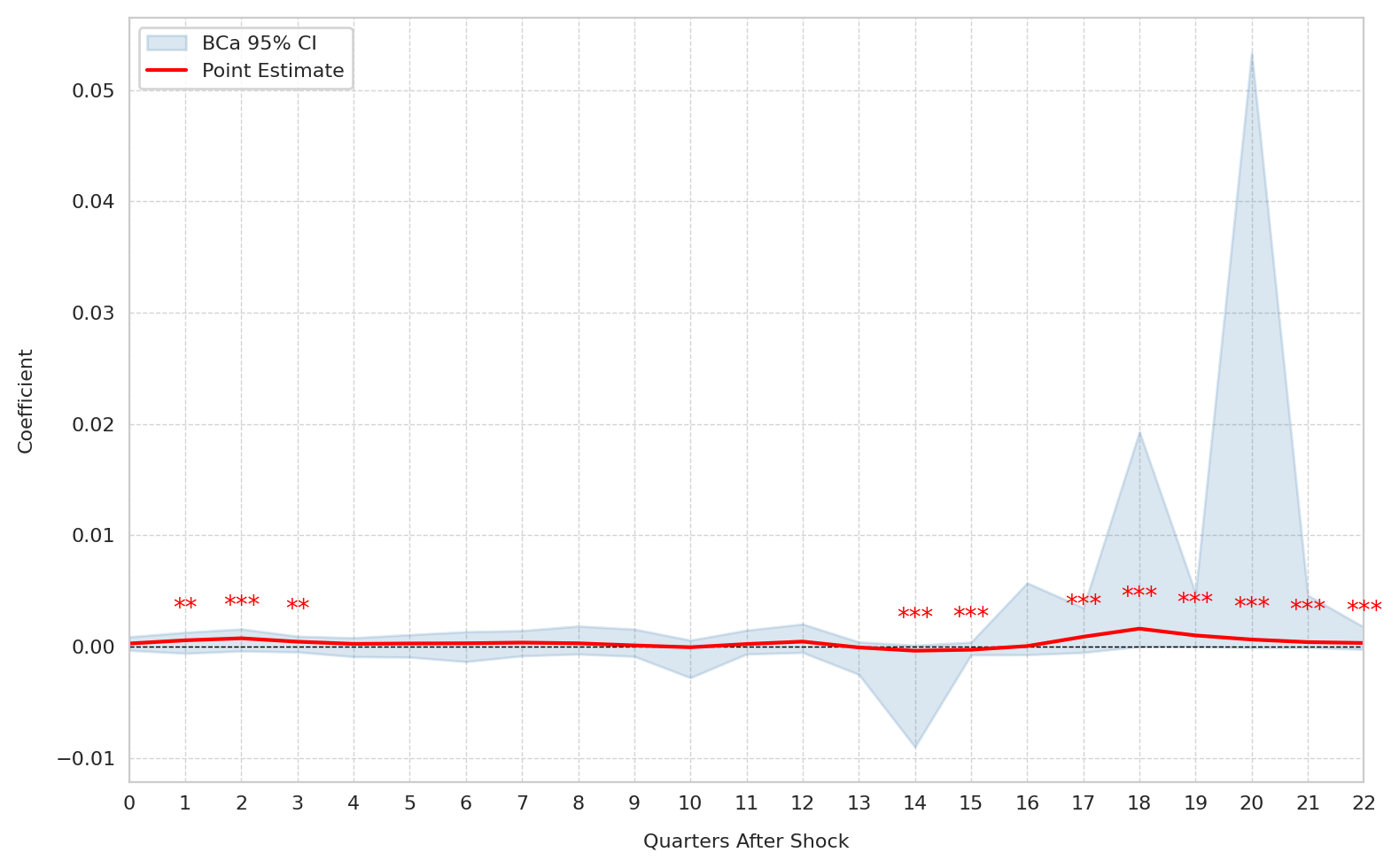}
    \caption{Impulse Response of Government Final Consumption}
    \label{fig:GC_response}
\end{minipage}
\hfill
\begin{minipage}[b]{0.48\linewidth}
    \centering
    \includegraphics[width=\linewidth]{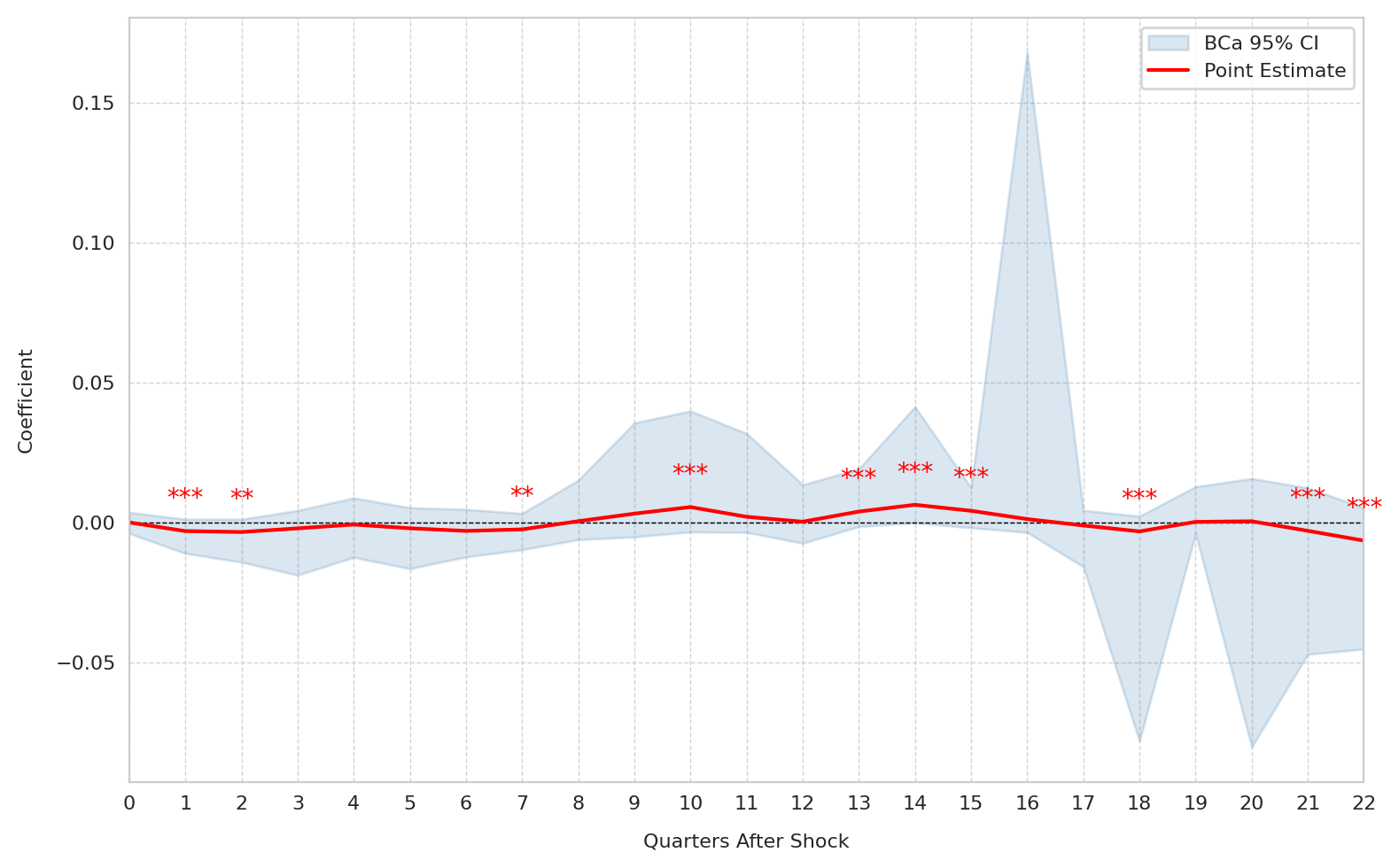}
    \caption{Impulse Response of Industry value added/GDP}
    \label{fig:lV}
\end{minipage}
\end{figure*}
From a growth-theoretic perspective, narrative-based reforms and structural changes have a lagged impact on productivity, typically requiring capital reallocation and organizational adjustment before improvements become visible. In our model, these processes unfold with minimal noise at firsthence relatively narrow bandsuntil a sufficiently strong signal (or simultaneous external shock) triggers a more abrupt revision in beliefs. When firms realign their investment and innovation decisions en masse, measured productivity can jump in ways that are large but also highly variable. As a result, a few data points in that window carry disproportionate weight, causing the standard errors to widen. Thus, the volatility in productivity's IRF aligns with the model's prediction that shifts from partial to widespread adoption of a policy narrative can yield transitory surges in uncertainty. It is also worth mentioning that although we include the COVID dummy in our specification, we still witness significantly negative response, this result can also be caused by the ``expectation shock" caused by the pandemic, and such impact cannot be fully captured by the dummy.

\textbf{Technology Expenditure \& Manufacturing Investment.} The IRF for Technology Expenditure (Figure 5) shows an overall positive and statistically significant response through much of the early and middle horizons, aligning with the core theoretical proposition that credible narratives trigger rapid fiscal prioritization toward innovation. However, the pronounced dip with a wide confidence interval around horizons 20–21 underscores a key tension in our model: once the Construct phase begins to wane (i.e., once the narrative is no longer ``new information") and external shocks or fiscal constraints come into play, technology-related spending may turn volatile. From the Bayesian perspective of our model, the private sector's belief revisions about future R\&D incentives can fluctuate sharply if either (i) signals about long-term commitment weaken or (ii) exogenous disruptive events arise (e.g., pandemic-induced budget reallocation or abrupt political shifts). The large negative swing and its sizeable uncertainty band thus likely reflect a period in which prior commitments to tech investment were challenged by unexpected macro shocks, in line with the notion that policy credibility is not immutable but contingent on sustained narrative reinforcement.

For Manufacturing Investment (Figure 6), one feature is the sporadic clustering of negative coefficients (particularly around horizons 12–18) amid occasionally wide confidence intervals. Early on, investment briefly dips significantly below zerosuggesting that while narratives may induce initial caution among manufacturers (perhaps interpreting the policy shift as favoring high-tech or services rather than traditional production), the effect largely remains close to zero. Eventually, as the narrative gains traction and the economy transitions toward the Commitment phase, the estimated responses become decidedly mixed, oscillating between modest positives and negatives. This pattern aligns with a more heterogeneous reallocation dynamic: firms in certain sub-sectors may respond to the new policy narrative by ramping up capital spending, while others might hold back in anticipation of regulatory changes or resource competition with more innovation-focused sectors. The net outcome is high variance in manufacturing investment decisionsand thus wide IRF confidence bands. 

\textbf{Government Final Consumption \& Industry value added/GDP.} Figures 7 and 8 underscore that government final consumption and industry value added do not appear to be primary transmission channels in our model of policy narratives. In Figure 7, government spending responds modestly in the earliest quarters, suggesting an initial budget reallocation consistent with the Construct phasebut remains close to zero (and often statistically indistinguishable from it) in subsequent horizons. One plausible explanation is that discretionary government consumption is relatively inflexible in our setting: once core budget priorities are set, narratives alone may not substantially alter near-term consumption patterns unless accompanied by legislative changes or exogenous events. The sizable spike in the standard error band near horizon 18–20 likely reflects sporadic fiscal shocks or other off-model policy interventions rather than the systematic working of narrative-driven expectations. Meanwhile, Figure 8 shows industry value added hovering near zero, with occasional significance in both directions, implying that the broad industrial sector is not the primary beneficiary (nor victim) of a narrative emphasizing innovation and technological upgrading. The irregular swings also point to nontrivial cross-industry reallocation under the surfacesome sectors might be crowding in while others are displacedaveraging out to relatively small net effects at the aggregate industry level. Thus, these muted (and sometimes erratic) responses are unsurprising in light of our framework, which posits that technology-based narratives would more directly impact R\&D expenditures and advanced manufacturing rather than the overall industry component.

\section{Implication and Discussion}
\pdfbookmark[1]{Implication and Discussion}{sec:Discussion}

Our analysis centers on a dynamic Bayesian game framework in which Innovation-Driven Development Strategy (IDDS) announcements act as narrative shocks, shaping agents' expectations and resource allocation. In contrast to standard ``one-shot" policy interventions, the IDDS relies on a sustained feedback loop between central government signals and local implementation efforts. Under this mechanism, the credibility of the policy commitment emerges only after repeated and consistent reinforcement of the original narrative, leading to the eventual institutionalization of industrial and fiscal directives. A crucial insight is that industrial policy unfolds in a more complex ecosystem than often acknowledged by traditional models. Our findings suggest that short-run market responses to IDDS announcements (e.g., spike in technology expenditure, delayed reaction in GDP) derive not from a single exogenous shock, but from a continuous narrative constructed by central authorities and subsequently validatedor, in some cases, underminedby local enforcement and budgetary frameworks. This endogenously evolving credibility underpins our NCC (Narratives–Construct–Commitment) approach, in which initial policy articulation (Narratives) triggers expectation anchoring (Construct), ultimately culminating in stable, self-enforcing growth drivers (Commitment) \footnote{We are working on improving the local projection section.}.

Our empirical findings aspire to inform the nascent Quality Productive Forces initiative, underscoring that government-led narratives create a guiding vision (the Construct stage) which must be gradually institutionalized through fiscal, industrial, and regulatory policies to become a robust pillar for long-run growth (Commitment). Because industrial policy operates in this layered environment, it should be viewed as an iterative process rather than a discrete event. Three policy considerations emerge. Firstly, the fidelity of policy commitment depends critically on local governments' consistency, continuity, and adaptability in executing central directives. Local authoritiesbeing responsible for real-world enactmentmust fully internalize the top-level design of IDDS-like narratives and ensure alignment between central objectives and local development paths, forming a closed policy loop from the center to the grassroots. Secondly, turnover in local governments can disrupt policy momentum. Hence, it is imperative to establish institutionalized mechanisms that ensure not only the inheritance of established policies but also their judicious refinement as circumstances evolve. Such frameworks safeguard against abrupt policy reversals and enable a smoother progression from narrative articulation to tangible, innovation-oriented outcomes. Thirdly, local governments must tailor the policy agenda to their distinct economic conditions and development needs. This includes identifying emerging industries, reallocating resources in line with new-quality productivity requirements, and enhancing the broader socio-economic returns of the reform agenda. In essence, a well-designed narrative can galvanize agents initially, but only consistent, transparent, and context-aware local enforcement cements it as a long-term driver.

While our current evidence supports the NCC framework and highlights the multifaceted role of policy narratives, several avenues remain for further exploration. We are finalizing the mathematical appendix and conducting numerical simulations to refine the theoretical model's parameterizations, evaluating how varying degrees of local government heterogeneity and budget constraints influence the Construct-to-Commitment transition. Additionally, we plan to expand our robustness analyses. 

We are also incorporating a dynamic reverse auction framework as a micro-level extension of the NCC model. In this setting, local governments periodically solicit bids from firms, providing policy support as ``rewards," while firms offer potential co-investments or productivity commitments. By treating each government-firm match as a repeated reverse auction with asymmetric information and evolving reputations, we can formally examine how localized policy competition and firm bidding strategies interact with the macro narrative signals. Preliminary analysis suggests that this dynamic mechanism helps clarify how long-term credibility forms at the granular level, complementing the broader Construct-to-Commitment perspective.

More broadly, our analysis speaks to the role of narrative authority vis-à-vis institutional capacity in transitional economies. It offers a novel lens for interpreting China's policy transmission dynamics under the New Quality Productive Forces initiative and contributes to the theoretical architecture of Chinese-style modernization. As the narrative ecosystem matures and data availability improves, our future research will extend this framework to disentangle how credibility, coordination, and commitment co-evolve across levels of governance and economic organization.

\section*{References}

\textbf{Acemoglu, Daron, Ufuk Akcigit, Harun Alp, Nicholas Bloom, and William Kerr.} 2019. ``Innovation, Reallocation, and Growth.“ \textit{American Economic Review} 109 (11): 3450--3491.

\textbf{Acemoglu, Daron, Simon Johnson, and James A. Robinson.} 2005. ``Institutions as a Fundamental Cause of Long-Run Growth.“ In \textit{Handbook of Economic Growth}, edited by Philippe Aghion and Steven N. Durlauf, 385--472. Amsterdam: Elsevier.

\textbf{Akerlof, George A.} 1970. ``The Market for ‘Lemons’: Quality Uncertainty and the Market Mechanism.“ \textit{Quarterly Journal of Economics} 84 (3): 488--500.

\textbf{Auerbach, Alan J., and Yuriy Gorodnichenko.} 2012. ``Measuring the Output Responses to Fiscal Policy.“ \textit{American Economic Journal: Economic Policy} 4 (2): 1--27.

\textbf{Baker, Scott R., Nicholas Bloom, and Steven J. Davis.} 2016. ``Measuring Economic Policy Uncertainty.“ \textit{Quarterly Journal of Economics} 131 (4): 1593--1636.

\textbf{Berger, Peter L., and Thomas Luckmann.} 1966. \textit{The Social Construction of Reality: A Treatise in the Sociology of Knowledge}. New York: Doubleday.

\textbf{Bernanke, Ben S.} 2020. ``The New Tools of Monetary Policy.“ \textit{American Economic Review} 110 (4): 943--983.

\textbf{Bernanke, Ben S., and Kenneth N. Kuttner.} 2005. ``What Explains the Stock Market’s Reaction to Federal Reserve Policy?“ \textit{Journal of Finance} 60 (3): 1221--1257.

\textbf{Entman, Robert M.} 1993. ``Framing: Toward Clarification of a Fractured Paradigm.“ \textit{Journal of Communication} 43 (4): 51--58.

\textbf{Evans, Charles L.} 2017. ``Odyssean and Delphic Guidance.“ \textit{Journal of Monetary Economics} 90: 62--77.

\textbf{Farhi, Emmanuel, and Iván Werning.} 2019. ``Monetary Policy, Bounded Rationality, and Incomplete Markets.“ \textit{American Economic Review} 109 (11): 3887--3928.

\textbf{Gabaix, Xavier.} 2020. ``Behavioral Inattention.“ In \textit{Handbook of Behavioral Economics}, edited by B. Douglas Bernheim, Stefano DellaVigna, and David Laibson, 261--343. Amsterdam: Elsevier.

\textbf{Gürkaynak, Refet S., Brian Sack, and Eric T. Swanson.} 2005. ``Do Actions Speak Louder Than Words? The Response of Asset Prices to Monetary Policy Actions and Statements.“ \textit{International Journal of Central Banking} 1 (1): 55--93.

\textbf{Hall, Peter A.} 1993. ``Policy Paradigms, Social Learning, and the State: The Case of Economic Policymaking in Britain.“ \textit{Comparative Politics} 25 (3): 275--296.

\textbf{Imbens, Guido W., and Thomas Lemieux.} 2008. ``Regression Discontinuity Designs: A Guide to Practice.“ \textit{Journal of Econometrics} 142 (2): 615--635.

\textbf{Jordà, Òscar.} 2005. ``Estimation and Inference of Impulse Responses by Local Projections.“ \textit{American Economic Review} 95 (1): 161--182.

\textbf{Kahneman, Daniel.} 1973. \textit{Attention and Effort}. Englewood Cliffs, NJ: Prentice-Hall.

\textbf{Kahneman, Daniel, and Amos Tversky.} 1979. ``Prospect Theory: An Analysis of Decision under Risk.“ \textit{Econometrica} 47 (2): 263--291.

\textbf{Kahneman, Daniel, and Amos Tversky.} 1981. ``The Framing of Decisions and the Psychology of Choice.“ \textit{Science} 211 (4481): 453--458.

\textbf{Kingdon, John W.} 1984. \textit{Agendas, Alternatives, and Public Policies}. Boston: Little, Brown.

\textbf{Kydland, Finn E., and Edward C. Prescott.} 1977. ``Rules Rather Than Discretion: The Inconsistency of Optimal Plans.“ \textit{Journal of Political Economy} 85 (3): 473--491.

\textbf{Lucas, Robert E.} 1988. ``On the Mechanics of Economic Development.“ \textit{Journal of Monetary Economics} 22 (1): 3--42.

\textbf{McCombs, Maxwell E., and Donald L. Shaw.} 1972. ``The Agenda-Setting Function of Mass Media.“ \textit{Public Opinion Quarterly} 36 (2): 176--187.

\textbf{Mertens, Karel, and Morten O. Ravn.} 2013. ``The Dynamic Effects of Personal and Corporate Income Tax Changes in the United States.“ \textit{American Economic Review} 103 (4): 1212--1247.

\textbf{Nakamura, Emi, and Jón Steinsson.} 2018. ``High-Frequency Identification of Monetary Non-Neutrality: The Information Effect.“ \textit{Quarterly Journal of Economics} 133 (3): 1283--1330.

\textbf{Newey, Whitney K., and Kenneth D. West.} 1987. ``A Simple, Positive Semi-definite, Heteroskedasticity and Autocorrelation Consistent Covariance Matrix.“ \textit{Econometrica} 55 (3): 703--708.

\textbf{Plagborg-Møller, Mikkel, and Christian K. Wolf.} 2021. ``Local Projections and VARs Estimate the Same Impulse Responses.“ \textit{Econometrica} 89 (2): 955--980.

\textbf{Ramey, Valerie A.} 2011. ``Identifying Government Spending Shocks: It’s All in the Timing.“ \textit{Quarterly Journal of Economics} 126 (1): 1--50.

\textbf{Ramey, Valerie A.} 2016. ``Macroeconomic Shocks and Their Propagation.“ In \textit{Handbook of Macroeconomics}, edited by John B. Taylor and Harald Uhlig, 71--162. Amsterdam: Elsevier.

\textbf{Romer, Christina D., and David H. Romer.} 2004. ``A New Measure of Monetary Shocks: Derivation and Implications.“ \textit{American Economic Review} 94 (4): 1055--1084.

\textbf{Romer, Paul M.} 1990. ``Endogenous Technological Change.“ \textit{Journal of Political Economy} 98 (5, part 2): S71--S102.

\textbf{Severin, Werner J., and James W. Tankard.} 2001. \textit{Communication Theories: Origins, Methods, and Uses in the Mass Media}. 5th ed. New York: Addison-Wesley Longman.

\textbf{Shiller, Robert J.} 2017. ``Narrative Economics.“ \textit{American Economic Review} 107 (4): 967--1004.

\textbf{Simon, Herbert A.} 1957. \textit{Models of Man: Social and Rational}. New York: Wiley.

\textbf{Smets, Frank, and Rafael Wouters.} 2007. ``Shocks and Frictions in US Business Cycles: A Bayesian DSGE Approach.“ \textit{American Economic Review} 97 (3): 586--606.

\textbf{Spence, Michael.} 1973. ``Job Market Signaling.“ \textit{Quarterly Journal of Economics} 87 (3): 355--374.

\textbf{Stiglitz, Joseph E.} 1975. ``The Theory of ‘Screening,’ Education, and the Distribution of Income.“ \textit{American Economic Review} 65 (3): 283--300.

\textbf{Stock, James H., and Mark W. Watson.} 2018. ``Identification and Estimation of Dynamic Causal Effects in Macroeconomics Using External Instruments.“ \textit{Economic Journal} 128 (610): 917--948.


\end{document}